\documentclass[journal]{IEEEtran}
\usepackage{amsfonts}
\usepackage{balance}
\usepackage{booktabs}
\usepackage{amsmath}
\usepackage{tabularx}
\usepackage{subfigure}
\usepackage{graphicx}
\usepackage{multirow}
\usepackage{rotating}
\usepackage{algorithm}
\usepackage{url}
\usepackage{amsthm}
\usepackage{algcompatible}
\usepackage{epstopdf}
\usepackage{makecell}
\usepackage{bm}
\usepackage{threeparttable}

\usepackage{color}

\begin{document}
\title{Broad Recommender System: An Efficient Nonlinear Collaborative Filtering Approach}

\author{Ling Huang,~\IEEEmembership{Member,~IEEE,}
        Can-Rong Guan, Zhen-Wei Huang, Yuefang Gao,\\ Chang-Dong~Wang,~\IEEEmembership{Senior Member,~IEEE},
        and C. L. Philip Chen,~\IEEEmembership{Fellow,~IEEE}
\thanks{This work was supported by NSFC (62276277 and 62106079) and Guangdong Basic and Applied Basic Research Foundation (2022B1515120059).}
\thanks{L. Huang, C.-R. Guan, Z.-W. Huang, and Y. Gao are with Department of Computer Science, South China Agricultural University, Guangzhou, China. E-mail: huanglinghl@hotmail.com, guancanrong@stu.scau.edu.cn, huangzhenwei@stu.scau.edu.cn, gaoyuefang@scau.edu.cn.}
\thanks{C.-D. Wang is with Department of Computer Science, Sun Yat-sen University, Guangzhou, China.
        E-mail: changdongwang@hotmail.com.}
\thanks{C. L. P. Chen is with Department of Computer Science, South China University of Technology, Guangzhou, China.
E-mail: philip.chen@ieee.org.}
}

\markboth{}%
{\MakeLowercase{\textit{Huang et al.}}: BroadCF}

\maketitle

\begin{abstract}
Recently, Deep Neural Networks (DNNs) have been largely utilized in Collaborative Filtering (CF) to produce more accurate recommendation results due to their ability of extracting the nonlinear relationships in the user-item pairs. However, the DNNs-based models usually encounter high computational complexity, \textit{i.e.}, consuming very long training time and storing huge amount of trainable parameters. To address these problems, we develop a novel broad recommender system named Broad Collaborative Filtering (BroadCF), which is an efficient nonlinear collaborative filtering approach. Instead of DNNs, Broad Learning System (BLS) is used as a mapping function to learn the nonlinear matching relationships in the user-item pairs, which can avoid the above issues while achieving very satisfactory rating prediction performance. {Contrary to DNNs, BLS is a shallow network that captures nonlinear relationships between input features simply and efficiently.} However, directly feeding the original rating data into BLS is not suitable due to the very large dimensionality of the original rating vector. To this end, a new preprocessing procedure is designed to generate user-item rating collaborative vector, which is a low-dimensional user-item input vector that can leverage quality judgments of the most similar users/items. Convincing experimental results on seven datasets have demonstrated the effectiveness of the BroadCF algorithm.

\end{abstract}

\IEEEpeerreviewmaketitle

\section{Introduction}
\label{sec:introduction}
\IEEEPARstart{I}{n} the era of information explosion, the rich choices of online services lead to an increasingly heavy role for recommender systems (RSs). The goal of RSs is to recommend the preferable items to users based on their latent preferences extracted from the historical behavioral data~\cite{DBLP:journals/tnn/WangZYXD21,DBLP:journals/dase/MaWZCL19, DBLP:conf/ijcai/HuDHL19,DBLP:journals/tetci/ChenZGW23,DBLP:conf/www/XuZLXSCZX19, DBLP:conf/recsys/BarkanKYK19, DBLP:journals/tnn/WangNL20,DBLP:conf/aaai/FuPWXL19, DBLP:conf/recsys/Ferraro19,DBLP:journals/tetci/WangCLLJX22,DBLP:journals/dase/ChenCHFLLZ20}. Collaborative filtering is one of the most classical recommendation algorithms, which has been widely used in many real-world tasks~\cite{DBLP:conf/ijcai/0001DWTTC18, DBLP:journals/sigir/HerlockerKBR17, DBLP:conf/adaptive/SchaferFHS07, DBLP:conf/www/ChaeKKC19,DBLP:journals/tkde/ShinKSX18,DBLP:journals/tcyb/ZhangSZCW20}.

\IEEEpubid{\begin{minipage}{\textwidth}\ \\[40pt] \centering
		Copyright \copyright 2024 IEEE.  Personal use of this material is permitted.  Permission from IEEE must be obtained for all other uses, in any current or future media, including reprinting/republishing this material for advertising or promotional purposes, creating new collective works, for resale or redistribution to servers or lists, or reuse of any copyrighted component of this work in other works.
\end{minipage}}

The traditional collaborative filtering is user-oriented or item-oriented, which relies mainly on the rating information of similar users or items to predict unknown ratings. Among various collaborative filtering techniques, Matrix Factorization (MF)~\cite{DBLP:conf/aaai/ChenLZZL18,DBLP:journals/tkde/VlachosDHVPA19,DBLP:journals/tkde/HeWJ19,DBLP:journals/tnn/HeTDHRC20} is the most popular approach, which operates by learning the latent representations of users/item in the common space. In the common space, the recommender system predicts the rating of the user-item pair based on their latent representations. However, in practical applications, rating data is prone to long-tail distribution~\cite{Hu:2016:LIP:3001595.2976737}, which can lead to the sparsity problem in the MF methods. To overcome the sparsity problem, several methods have been developed~\cite{DBLP:journals/csur/ShiLH14}, \textit{e.g.}, adding social relations to MF or using invisible feedback~\cite{svd++}.

Deep Neural Networks (DNNs) have been rapidly developed recently, which have shown satisfactory results in different fields such as computer vision and natural language processing. DNNs have also been introduced into the field of RSs~\cite{DBLP:journals/tkde/ShiHZY19,xue2017deep,He:2017:NCF:3038912.3052569, DBLP:conf/sigir/LiuW0WJWX19,DBLP:journals/tkde/WangXHZHL22,DBLP:journals/tcyb/ChenTS22}. In the traditional matrix decomposition, the relationships in the user-item pairs are usually assumed to be linear, which is however not true in real-world scenarios. To better learn nonlinear matching relationships in the user-item pairs, Xue \textit{et al.} propose Deep Matrix Factorization (DMF)~\cite{xue2017deep} that leverages a dual-path neural network instead of the linear embedding adopted in the conventional matrix decomposition. DNNs are able to approximate any function, so they are well suitable for learning complex matching functions. For example, He et al. propose the Neural Collaborative Filtering (NCF) framework~\cite{He:2017:NCF:3038912.3052569}, which takes as input the concatenation of user-item embeddings and uses Multi-Layer Perceptron (MLP) for prediction. Deng \textit{et al.} propose DeepCF~\cite{DeepCF} which is a unified deep collaborative filtering framework integrating matching function learning and representation learning. However, the above DNNs-based models require a large number of training epochs that are computationally expensive, and therefore cannot be applied to large-scale data and are very time consuming. As shown in our experiments, due to the out-of-memory error, most DNNs-based models fail to generate results when dealing with an Amazon dataset containing 429622 users, 23966 items, and 583933 ratings on a server with an Intel Core i9-10900 CPU, GeForce RTX 3090, and 256GB of RAM. And even for the relatively small datasets that can generate results, most DNNs-based models are time-consuming. In~\cite{DBLP:journals/tkde/WangXHZHL22}, a BP Neural Network with Attention Mechanism (BPAM) has been developed to mitigate this problem by combining attention mechanism with the BP network. However, the gradient-based optimization still consumes long time.

To address the aforementioned problems, this paper proposes a novel broad recommender system, called Broad Collaborative Filtering (BroadCF), which is an efficient nonlinear collaborative filtering approach. Different from the DNNs-based methods that feed the user-item rating data into DNNs, the proposed BroadCF method feeds the user-item rating data into Broad Learning System (BLS)~\cite{DBLP:journals/tnn/ChenLS19}. {Broad Learning {System} was first published in the Computational Intelligence Society~\cite{DBLP:journals/tnn/ChenLS19}, {which then becomes a very important  research topic in this field~\cite{DBLP:journals/tfs/HuangVWQZC23,DBLP:journals/tfs/BaiZWZZ22,DBLP:journals/tai/DaiZWCW21,9938406}.}} As shown in the literature~\cite{DBLP:journals/tnn/ChenLS19,DBLP:journals/tcyb/GongZCL22,DBLP:journals/tetci/WangCCW23}, BLS is also able to approximate any function, and its general approximation capability has been theoretically analyzed at the beginning of its design and proved by rigorous theoretical proof. In particular, it can map the data points to a discriminative space by using random hidden layer weights via any continuous probability distribution~\cite{DBLP:journals/tnn/ChenLS19,DBLP:journals/tcyb/GongZCL22,DBLP:journals/tetci/XieVCW22}. This random mechanism provides a fast training process for BLS. That is, it only needs to train the weights from the hidden layer to the output layer by a pseudo-inverse algorithm. Therefore, BroadCF does not need to consume so much training time compared with the DNNs-based models, and is more suitable for larger datasets due to the small number of stored parameters.

However, it is not a trivial task to input the user-item rating data into BLS. The DNNs-based methods take the original user/item rating vectors (or their direct concatenation) as input, which is suitable because the DNNs-based methods learn the latent factor vectors by stacking several hidden layers, the dimensions of which usually decrease monotonously from input, hidden to output. Different from the DNNs-based methods, BLS needs to broaden the original input data into the mapped features and enhanced features with several-order magnitude larger dimensions. Therefore, it is not suitable to take the original user/item rating vectors (or their direct concatenation) as input for BLS. Inspired by the previous work \cite{DBLP:journals/tkde/WangXHZHL22}, this paper proposes a new preprocessing procedure to generate user-item rating collaborative vector, which is a low-dimensional user-item input data that can leverage quality judgments of the most similar users/items.

The contributions of this work are concluded as follows.
\begin{itemize}
	\item{A new broad recommender system called BroadCF is proposed to overcome the problems of long training time and huge storage requirements in the DNNs-based models.}
	\item{A novel data preprocessing procedure is designed to convert the original user-item rating data into the form of (Rating collaborative vector, Rating) as input and output of BroadCF.}
	\item{Extensive experiments are performed on seven datasets to confirm the effectiveness of the model. The results show that BroadCF performs significantly better than other state-of-the-art algorithms and reduces training time consumption and storage costs.}
\end{itemize}

The remainder of the paper is organized as follows. We will review the related work in Section~\ref{sec:RelatedWork}. In Section~\ref{sec:Problem Statement and Preliminaries}, we will introduce the problem statement and preliminaries. We will detail the BroadCF method in Section~\ref{sec:BroadCF}. In Section~\ref{sec:Experiments}, the experimental results will be reported with convincing analysis. We will draw the conclusion {and present the future work} in Section~\ref{sec:Conclusion}.

\begin{table*}[!t]
	\caption{Main ideas and specificities of the {related} work and BroadCF.}
	\label{tab:related work}
	\vskip -0.1in
	\centering
        \begin{threeparttable}
		\begin{tabular}{@{}c@{}rrrrrrrrr@{}}
			\hline
			\multicolumn{1}{c}{\textbf{}} & \multicolumn{1}{c}{\textbf{PMF}~{\cite{mnih2008probabilistic}}} & \multicolumn{1}{c}{\textbf{NeuMF}~{\cite{He:2017:NCF:3038912.3052569}}} & \multicolumn{1}{c}{\textbf{DMF}~{\cite{xue2017deep}}}  & \multicolumn{1}{c}{\textbf{DeepCF}~{\cite{DeepCF}}} & \multicolumn{1}{c}{\textbf{BPAM}~{\cite{DBLP:journals/tkde/WangXHZHL22}}} & \multicolumn{1}{c}{\textbf{SHT}~{\cite{DBLP:conf/kdd/Xia0Z22}}} & \multicolumn{1}{c}{\textbf{NGCF}~{\cite{DBLP:conf/sigir/Wang0WFC19}}} & \multicolumn{1}{c}{\textbf{LigthGCN}~{\cite{DBLP:conf/sigir/0001DWLZ020}}} & \multicolumn{1}{c@{}}{\textbf{{BroadCF}}}\\
			\hline
			\textbf{{Main idea}} & MF & MF & MF & CF & MF, CF & CF & CF & CF & MF, CF \\
   			\textbf{Nonlinear} & $\times$ & $\checkmark$ & $\checkmark$ & $\checkmark$ & $\checkmark$ & $\checkmark$ & $\checkmark$ & $\checkmark$ & $\checkmark$ \\
			{\textbf{Training time}} & {More} & {Medium} & {More} & {More} & {More} & {Medium} & {Medium} & {Medium} & {Less}\\
			\hline
	\end{tabular}
    \begin{tablenotes}
     \item[1] MF denotes Matrix Factorization, {and} CF denotes Collaborative Filtering.
     \item[2] Nonlinear denotes whether the nonlinear relationship between users and items is learned.
     \item[3] {More, Medium, and Less denote the degree of training time consumption. More denotes that the training time on the ml-la dataset is more than 1000s, and Less denotes that the training time on the ml-la dataset is less than 100s.}
   \end{tablenotes}
  \end{threeparttable}
  \vskip -0.1in
\end{table*}

\section{Related Work}
\label{sec:RelatedWork}
\subsection{Shallow Learning based Collaborative Filtering Methods}
Early collaborative filtering in its original form is user-oriented or item-oriented~\cite{DBLP:journals/internet/LindenSY03}, which relies primarily on the rating information of similar users/items to predict unknown ratings. Many efforts have been made in developing variants of CF from different perspectives~\cite{mnih2008probabilistic,DBLP:journals/tcyb/WangDLY19,4470228,5474122,PATRA2015163}. {Probabilistic Matrix Factorization (PMF) considers	only latent factors by using matrix decomposition to capture the linear relationships in the user-item pairs~\cite{mnih2008probabilistic}.} In~\cite{4470228}, Bell and Koren propose to learn the interpolation weights from the rating data as the global solution, which leads to an optimization problem of CF and improves the rating prediction performance. In~\cite{svd++}, Koren {proposes} to extend the SVD-based model to the SVD++ model by utilizing the additional item factors to model the similarity among items.
Another early attempt is~\cite{5474122}, in which the temporal information is considered for improving the accuracy. Instead of relying on the co-rated items, Patra et al. propose to utilize all rating information for searching useful nearest users of the target user from the rating matrix~\cite{PATRA2015163}. In order to recommend new and niche items to users, based on the observation of the Rogers' innovation adoption curve~\cite{krueger2006impact}, Wang et al. propose an innovator-based collaborative filtering algorithm by designing a new concept called innovators~\cite{DBLP:journals/tcyb/WangDLY19}, which is able to balance serendipity and accuracy. For more literature reviews of the shallow learning based collaborative filtering methods, please refer to the survey papers~\cite{DBLP:journals/csur/ShiLH14,DBLP:journals/tsc/ZhengLTXL22}. Despite the wide applications, the conventional shallow learning based CF methods assume the linear relationships in the user-item pairs, which is however not true in the real-world applications.

\subsection{Deep Learning based Collaborative Filtering Methods}

Recently, DNNs have been widely adopted in CF to learn the complex mapping relationships in the user-item pairs due to their ability to approximate any continuous function. For example, in~\cite{xue2017deep}, Xue et al. develop a new matrix decomposition model based on a neural network structure. It maps users and items into a shared low-dimension space via a nonlinear projection. He et al. develop a neural network structure to capture the latent features of users/items, and design a general framework called Neural Collaborative Filtering (NCF)~\cite{He:2017:NCF:3038912.3052569}{, which contains some very well-known approaches such as Neural Matrix Factorization (NeuMF)}. In~\cite{DeepCF}, Deng et al. propose a unified deep CF framework integrating matching function learning and  representation learning. In~\cite{DBLP:journals/tois/XueHWXLH19}, Xue et al. propose a deep item-based CF method, which takes into account of the higher-order relationship among items. In~\cite{DBLP:journals/tcyb/FuQYLL19}, Fu et al. propose a multi-view feed-forward neural network which considers the historical information of both user and item. In~\cite{DBLP:journals/tnn/HanZXZZYZ20}, Han et al. develop a DNNs-based recommender framework with an attention factor, which can learn the adaptive representations of users.
In~\cite{DBLP:journals/tcyb/LiuYL23}, Liu et al. propose a dual-message propagation mechanism for graph collaborative filtering, which can model preferences and similarities for recommendations.
In~\cite{DBLP:journals/tcyb/ZhongHWLY22}, Zhong et al. propose a cross-domain recommendation method, which utilizes autoencoder, MLP, and self-attention to extract and integrate the latent factors from different domains. In~\cite{DBLP:conf/kdd/Xia0Z22}, {Xia et al. propose Self-supervised Hypergraph Transformer (SHT), which introduces the hypergraph transformer network} for collaborative filtering recommendation via self-supervised learning. For more literature reviews of the DNNs-based CF methods, please refer to the survey paper~\cite{SurveyDeepRec:22}.
However, due to the large amount of trainable parameters and complex structures involved, the DNNs-based models usually encounter high computational complexity, \textit{i.e.}, consuming very long training time and storing huge amount of trainable parameters. {{Neural Graph
Collaborative Filtering (NGCF)~\cite{DBLP:conf/sigir/Wang0WFC19}} is a collaborative filtering recommendation method based on graph neural networks. It utilizes graph neural networks to extract higher-order interaction information between users and items to improve {the} recommendation performance. LightGCN~\cite{DBLP:conf/sigir/0001DWLZ020} is a lightweight graph convolution collaborative filtering recommendation method. It uses only the neighbor aggregation part of {Graph Convolutional Network (GCN)} and weights the higher-order embeddings of users and items learned on all layers to get the predicted rating. {Incremental GCN (IGCN)~\cite{DBLP:conf/cikm/XiaLGLZG21}} is an incremental GNN-based recommendation model {that updates} the time-aware properties of users and items. {The main ideas and specificities of the related work and BroadCF} are summarized in Table~\ref{tab:related work}.}

In this paper, to overcome the above problems, we propose a novel neural network based recommender system called Broad Collaborative Filtering (BroadCF). Instead of DNNs, Broad Learning System (BLS) is used as a mapping function to learn the nonlinear matching relationships in the user-item pairs, which can avoid the above issues while achieving very satisfactory rating prediction performance.

\section{Problem Statement and Preliminaries}
\label{sec:Problem Statement and Preliminaries}

In this section, we will present the problem statement and preliminaries. Table~\ref{table:notations} summarizes the major notations used throughout this paper.

\begin{table}[!t]
	\caption{The major notations.}
	\label{table:notations}
	\vskip -0.1in
	\centering{
		\resizebox{\linewidth}{!}{
			\begin{tabular}{@{}c@{}|l@{}}
				\hline
				$\mathcal{U}$, $\mathcal{V}$                	&User set, item set\\
				\hline
				{$\mathbf{R}  \in \mathbb{R}^{\left| \mathcal{U} \right|  \times \left| \mathcal{V} \right| } $} & User-item rating matrix\\
				\hline
				$r_{u,v}$			     	&Rating of user $u$ to item $v$\\
				\hline
				$\hat{r}_{u,v}$		&Predicted rating of user $u$ to item $v$\\
				\hline
				$k$ & Number of the nearest users of each user $u$ \\
				\hline
				\multirow{2}*{$\bm{\pi}^u\in\mathbb{R}^{1\times k}$} & Index vector storing the indexes of $k$ nearest users (KNU)\\
& of user $u$\\
				\hline
				$\bm{\pi}^u[i]$&Index of the $i$-th nearest user of user $u$\\
				\hline
				\multirow{2}*{$  \mathbf{p}^{{u}}_{v}  \in\mathbb{R}^{1\times k}$}  & KNU rating vector of user $u$ on item $v$, \textit{i.e.} the ratings\\& of the $k$ nearest users of user $u$ on item $v$\\
				\hline
				{$\mathbf{\bar{p}}^{{u}}_{v} \in\mathbb{R}^{1\times k}$} 	&{Post-processed KNU rating vector of user $u$ on item $v$}\\
				\hline
				$l$ & Number of the nearest items of each item $v$ \\
				\hline
				\multirow{2}*{$\bm{\eta}^v\in\mathbb{R}^{1\times l}$} & Index vector storing the indexes of $l$ nearest items (LNI)\\
& of item $v$\\
				\hline
				$\bm{\eta}^v[i]$&Index of the $i$-th nearest item of item $v$\\
				\hline
				\multirow{2}*{$\mathbf{q}^{{u}}_{v} \in\mathbb{R}^{1\times l}$}  & LNI rating vector of item $v$ from user $u$, \textit{i.e.} the ratings\\& of user $u$ to the $l$ nearest items of item $v$\\
				\hline
				{$\mathbf{\bar{q}}^{{u}}_{v} \in\mathbb{R}^{1\times l}$} 	&{Post-processed LNI rating vector of item $v$ from user $u$}\\
				\hline
				\multirow{2}*{$\mathbf{x}^{{u}}_{v}\in\mathbb{R}^{1\times (k+l)}$}& User-item rating collaborative vector for the given \\& user-item pair composed of user $ u $ and item $v$ \\
				\hline
				$\mathcal{D}$ & Training set of the proposed BroadCF algorithm\\
				\hline
				$\mathbf{X}\in\mathbb{R}^{|\mathcal{D}|\times (k+l)}$&Training input matrix of BLS\\
				\hline
				$\mathbf{y}\in\mathbb{R}^{|\mathcal{D}|\times 1}$&Training output vector of BLS\\
				\hline
				$d_z$ & Dimension of each mapped feature group\\
				\hline
				$\mathbf{Z}_i\in\mathbb{R}^{|\mathcal{D}|\times d_z}$ &The $i$-th mapped feature matrix\\
				\hline
				$n$& Number of the mapped feature groups\\
				\hline
				$\phi_i$& The $i$-th nonlinear feature mapping function\\
				\hline
				$d_h$ & Dimension of each enhanced feature group\\
				\hline
				$\mathbf{H}_i\in\mathbb{R}^{|\mathcal{D}|\times d_h}$ &The $i$-th enhanced feature matrix\\
				\hline
				$m$& Number of the enhanced feature groups\\
				\hline
				$\xi_i$& The $i$-th nonlinear feature enhancement function\\
				\hline
				$\mathbf{w}\in\mathbb{R}^{(nd_z+md_h)\times 1}$ & Learnable weight vector of BLS\\
				\hline
			\end{tabular}
		}
	}
 \vskip -0.1in
\end{table}

\subsection{Problem Statement}

Let $\mathcal{U}$ and $\mathcal{V}$ denote the user set and item set respectively. Following~\cite{DeepCF}, based on the users' ratings on items, we can obtain a user-item rating matrix $\mathbf{R}  \in \mathbb{R}^{\left| \mathcal{U} \right|  \times \left| \mathcal{V} \right| } $ as follows,
\begin{equation}
	{\mathbf{R}}_{u,v} =
	\begin{cases}
		r_{u,v}, &\mbox{if user } u \mbox{ has made rating on item } v\\
		0, &\mbox{otherwise}
	\end{cases}
	\label{RM}
\end{equation}
where $ r_{u,v} $ represents the rating score of user $ u $ {on} item $v$. Notice that, in most of the recommendation tasks, the user-item ratings are integers, \textit{e.g.}, $\{1,2,3,4,5\}$.

{The goal of RSs is} to predict the missing entries in the rating matrix $ \mathbf{R}  $. In the model-based methods, the predicted rating score $\hat{r}_{u,v}$ of user $u$ on item $v$ can be obtained as follows,
	\begin{equation}
		\label{eq:rf}
		\hat{r}_{u,v}= f(u,v| \Theta)
	\end{equation}
where $ f $ is a mapping function which outputs the predicted rating score of user $u$ on item $v$ by taking as input the rating information of neighbor users or items, and $ \Theta $ denotes the model parameters~\cite{DeepCF}.

In DNNs-based CF, $f$ inherits the capability of learning nonlinear representations from DNNs, and thus it is able to learn the nonlinear matching relationships in the user-item pairs. However, due to the large number of trainable parameters, the complex structure, and the continuous iterative training process, the DNNs-based models often suffer from high computational complexity, \textit{i.e.}, consuming very long training time and storing huge amount of trainable parameters. A recent work called BPAM~\cite{DBLP:journals/tkde/WangXHZHL22} has shed light on mitigating the problem of training time consumption by combining attention mechanisms with the BP networks. However, the gradient-based optimization still takes some time. In this paper, instead of DNNs, a lightweight neural network called Broad Learning System (BLS)~\cite{DBLP:journals/tnn/ChenLS19} is used as a mapping function to learn the nonlinear matching relationships in the user-item pairs, which can avoid the above issues while achieving very satisfactory rating prediction performance.

\begin{figure*}[!t]
	\centering
	\includegraphics[width=1.0\linewidth]{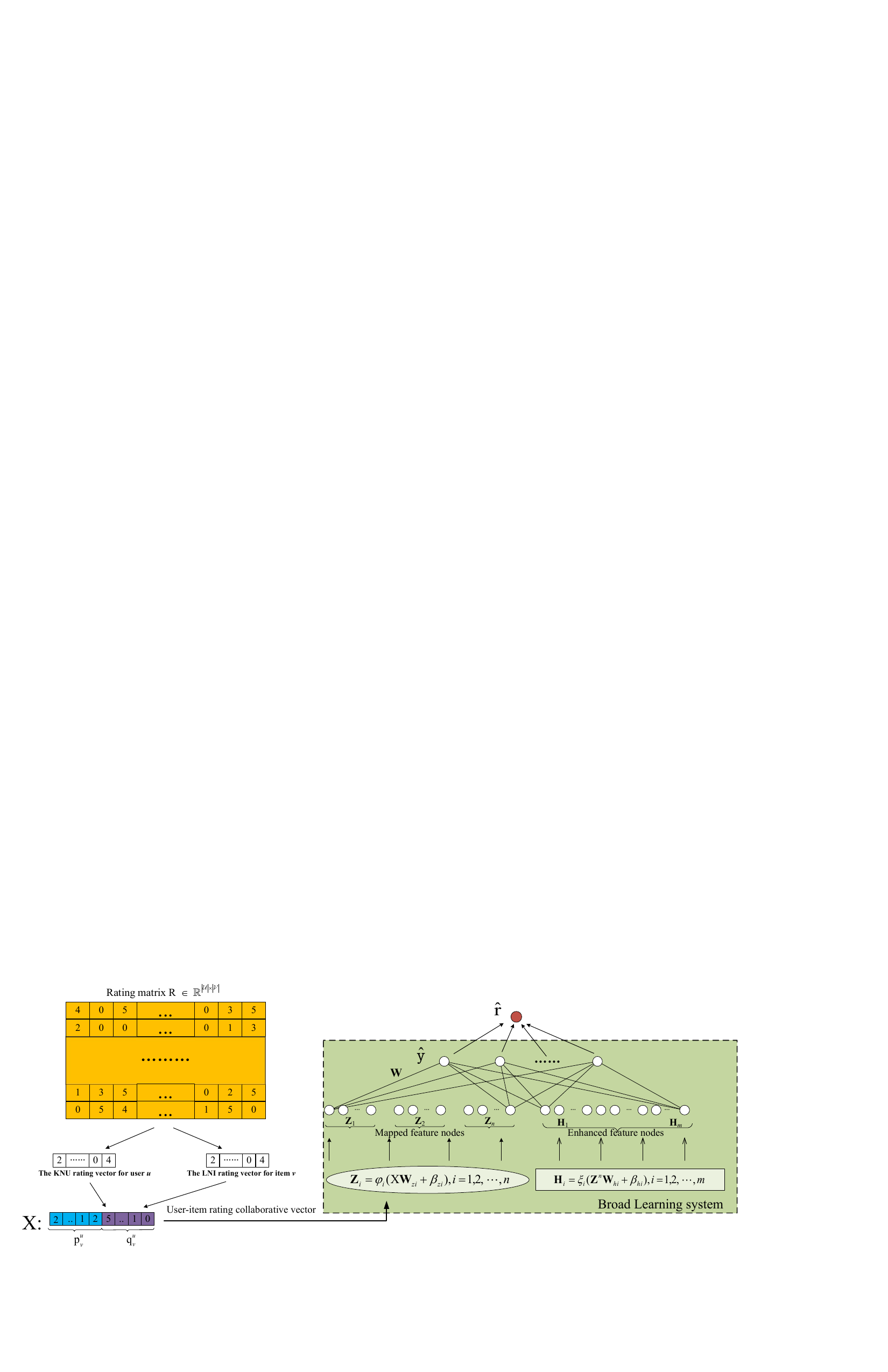}
	\caption{Illustration of the proposed BroadCF algorithm.}
	\label{figure:process}
\end{figure*}

\subsection{Preliminaries}
BLS is a lightweight neural network that can approximate any function~\cite{DBLP:journals/tnn/ChenLS19}, which has been widely used in many different applications~\cite{DBLP:journals/tcyb/ShengLZMC21,DBLP:journals/tcyb/DuVC21,Huang_SGP-FBLS:TFS23,Wang_BASS_BLS:TNNLS23,Cao_MSBLS_TCYB23,Xia_FTBLS:TII23}. It is derived from the Random Vector Functional Linked Neural Network (RVFLNN)~\cite{DBLP:journals/computer/PaoT92}. In BLS, the original data are firstly mapped into the mapped features by random weights, which are stored in the mapped feature nodes. Next, the mapped features are further mapped by random weights to obtain the enhanced features, which are used for extensive scaling. Finally, the parameter normalization optimization is solved using the ridge regression approximation to learn the final network weights.

The advantage of BLS is that it can use random hidden layer weights to map the original data points to a discriminative feature space tensed by these vectors. In the hidden layer, the parameters are randomly generated in accordance with any continuous probability distribution, the rationale of which has been theoretically proven. This random mechanism provides a fast training process for BLS. It solves the serious training time consuming problem suffered by the DNNs-based models. Specifically, BLS can be updated incrementally when adding {new} samples and hidden nodes, without the need of rebuilding the entire network from scratch. As a result, BLS requires a small amount of parameters to be stored, which can solve the problem of huge storage requirements arising from the DNNs-based models. Most importantly, in BLS, the feature layer is directly connected to the output layer, and the structure is very simple but effective. Therefore, even without a large amount of samples in RSs, the prediction performance can be ensured.

\section{Broad Collaborative Filtering}
\label{sec:BroadCF}

In this section, we will describe the proposed Broad Collaborative Filtering (BroadCF) method in detail. First of all, we will describe how to transform the original user-item rating data into the form of (Rating collaborative vector, Rating) for every user-item pair, which will be taken as input and output to BLS respectively. Then, we will describe the broad learning part. Finally, we will describe the rating prediction procedure and summarize the {proposed} BroadCF method. For clarity purpose, Figure~\ref{figure:process} illustrates the major procedure of the BroadCF method.

\subsection{Data Preprocessing}

From Eq. (\ref{eq:rf}), for each user-item pair $(u,v)$, the model-based RSs aim to establish a mapping function from the user-item input data to the user-item rating. A simple type of user-item input data is the direct concatenation of the rating vectors ${\mathbf{R}}_{u,:}$ and ${\mathbf{R}}_{:,v}$ of user $u$ and item $v$ respectively. It has been commonly adopted in the DNNs-based methods, \textit{i.e.}, the DNNs-based methods learn the latent factor vectors from the original user/item rating vectors. However, it does not work well in the case of BLS, as will be explained below. The DNNs-based methods stack several hidden layers with decreasing layer sizes, \textit{e.g.}, from the $10^4$-dimensional input to the first hidden layer of $10^3$ dimension, and further to the second hidden layer of $10^2$ dimension, etc. On the contrary, as will be shown later, BLS requires broadening the original input features into the mapped features and the enhanced features, \textit{e.g.}, from the $10^4$-dimensional input to the mapped features of $10^5$ dimension, and further to the enhanced features of $10^6$ dimension, resulting in several-order magnitude larger dimension. In addition, directly utilizing extremely sparse user rating {vectors} and item rating {vectors} may cause the overfitting issue. Inspired by the previous work~\cite{DBLP:journals/tkde/WangXHZHL22}, a new preprocessing procedure is designed to generate user-item rating collaborative vector, which is a low-dimensional user-item input vector that can leverage quality judgments of the most similar users/items.

\subsubsection{KNU Rating Vector}

First of all, for each user $u$, a set of $k$ nearest users (KNU) is searched by computing the cosine similarity between the rating vector of the target user $u$ and those of the other users. And
their indexes are stored in the KNU index vector $\bm{\pi}^u\in\mathbb{R}^{1\times k}$, \textit{i.e.}, the $i$-th nearest user of the target user $u$ can be expressed as $\bm{\pi}^u[i], \forall i= 1,\cdots, k$. Based on the KNU index vector $\bm{\pi}^u$ and the rating matrix $\mathbf{R}$, for the user-item pair $(u,v)$, we can obtain the KNU rating vector ${\mathbf{p}}_{v}^u \in \mathbb{R}^{1 \times k}$ of user $u$ on item $v$ with each entry ${\mathbf{p}}_{v}^u[i]$ being defined as follows,
\begin{eqnarray}
	\begin{aligned}
		\mathbf{p}_{v}^u[i] = \mathbf{R}_{\bm{\pi}^u[i],v}~~~
		\forall i= 1,\cdots,{k}.
	\end{aligned}
\end{eqnarray}
That is, ${\mathbf{p}}_{v}^u[i]$ is the rating value of user $\bm{\pi}^u[i]$ to item $v$.

In the above procedure, it is possible that some entries ${\mathbf{p}}_{v}^u[i]$ are zeros, \textit{i.e.}, user $\bm{\pi}^u[i]$ did not rate item $v$. In~\cite{DBLP:journals/tkde/WangXHZHL22}, the mean value of the ratings of user $\bm{\pi}^u[i]$ is used to fill the missing rating, which however, treats all the unrated items of user $\bm{\pi}^u[i]$ equally while failing to utilize the rating of other similar users to item $v$. To this end, a more reasonable strategy is designed as follows,
\begin{eqnarray}
	\label{eq:barp}
	\begin{aligned}
		&\mathbf{\bar{p}}_{v}^u[i] = \\
&\begin{cases}
            \mathbf{p}_{v}^u[i] & \text{~~~if~}\mathbf{p}_{v}^u[i]\neq 0\\
			\mathbf{R}_{u^*,v} & \text{~~~if~} \mathbf{p}_{v}^u[i]=0 \& \text{sim}(u^*,\bm{\pi}^u[i])\geq\theta\\
			\text{mean}(\mathbf{R}_{{\bm{\pi}^u[i]},:})& \text{~~~if~} \mathbf{p}_{v}^u[i]=0 \& \text{sim}(u^*,\bm{\pi}^u[i])<\theta
		\end{cases}
	\end{aligned}
\end{eqnarray}
where $u^*$ denotes the nearest user of user $\bm{\pi}^u[i]$ who has rated item $v$, $\text{sim}(u^*,\bm{\pi}^u[i])\geq\theta$ indicates that the similarity between user $u^*$ and user $\bm{\pi}^u[i]$ is no smaller than the threshold $\theta$, and $\text{mean}(\mathbf{R}_{{\bm{\pi}^u[i]},:})$ denotes the mean value of the ratings of user $\bm{\pi}^u[i]$. That is, when user $\bm{\pi}^u[i]$ did not rate item $v$ (\textit{i.e.} $\mathbf{p}_{v}^u[i]= 0$), we first search the most similar user of user $\bm{\pi}^u[i]$ who has already rated item $v$, denoted as user $u^*$. If the similarity between user $u^*$ and user $\bm{\pi}^u[i]$ is no smaller than the threshold $\theta$, we use the rating of user $u^*$ on item $v$ to fill the missing rating of user $\bm{\pi}^u[i]$ on item $v$. Otherwise, we use the mean value of the ratings of user $\bm{\pi}^u[i]$ to fill the missing rating of user $\bm{\pi}^u[i]$ on item $v$.

\subsubsection{LNI Rating Vector}

Similarly, for each item $v$, a set of $l$ nearest items (LNI) is searched by computing the cosine similarity between the rating vectors of the target item $v$ and the other items. And their indexes are stored in the LNI index vector $\bm{\eta}^v\in\mathbb{R}^{1\times l}$, \textit{i.e.}, the $i$-th nearest item of item $v$ can be expressed as $\bm{\eta}^v[i], \forall i= 1,\cdots , l$.
Based on the LNI index vector $\bm{\eta}^v$ and the rating matrix $\mathbf{R}$, for the user-item pair $(u,v)$, we can obtain the LNI rating vector $\mathbf{q}_{v}^u \in \mathbb{R}^{1 \times l}$ of user $u$ on item $v$ with each entry $\mathbf{q}_{v}^u[i]$ being defined as follows,
\begin{eqnarray}
	\begin{aligned}
		\mathbf{q}_{v}^u[i] = \mathbf{R}_{u,\bm{\eta}^v[i]}~~~
		\forall i= 1,\cdots,l.
	\end{aligned}
\end{eqnarray}
That is,\ $\mathbf{q}_{v}^u[i]$ denotes the rating value of user $u$ to item $\bm{\eta}^v[i]$.

Like before, the following strategy is used to fill the missing entries
\begin{eqnarray}
	\label{eq:barq}
	\begin{aligned}
		&\mathbf{\bar{q}}_{v}^u[i] = \\
&\begin{cases}
            \mathbf{q}_{v}^u[i] & \text{~~~if~}\mathbf{q}_{v}^u[i]\neq 0\\
			\mathbf{R}_{u,v^*} & \text{~~~if~} \mathbf{q}_{v}^u[i]=0\&\text{sim}(v^*,\bm{\eta}^v[i])\geq\theta\\
			\text{mean}(\mathbf{R}_{:,\bm{\eta}^v[i]})&\text{~~~if~} \mathbf{q}_{v}^u[i]=0\&\text{sim}(v^*,\bm{\eta}^v[i])<\theta
		\end{cases}
	\end{aligned}
\end{eqnarray}
where $v^*$ denotes the nearest item of item $\bm{\eta}^v[i]$ rated by user $u$, $\text{sim}(v^*,\bm{\eta}^v[i])\geq\theta$ indicates that the similarity between item $v^*$ and item $\bm{\eta}^v[i]$ is no smaller than the threshold $\theta$, and $\text{mean}(\mathbf{R}_{:,\bm{\eta}^v[i]})$ denotes the mean value of the ratings of item $\bm{\eta}^v[i]$.

\subsubsection{User-Item Rating Collaborative Vector}

After obtaining $\bar{\mathbf{p}}_{v}^u$ and $\bar{\mathbf{q}}_{v}^u$, a user-item rating collaborative vector is obtained by concatenation, \textit{i.e.},
\begin{equation}
	\label{eq:x_{v}^u}
	\mathbf{x}_{v}^u =         \left[
	\bar{\mathbf{p}}_{v}^u|\bar{\mathbf{q}}_{v}^u
	\right] \in \mathbb{R}^{1\times (k+l)}.
\end{equation}
Note that the dimension of the input data $\mathbf{x}_{v}^u$, \textit{i.e.} $k+l$, is much smaller than that of the direct concatenation of the rating vectors $\mathbf{R}_{u,:}$ and $\mathbf{R}_{:,v}$ of user $u$ and item $v$, \textit{i.e.} $|\mathcal{V}|+|\mathcal{U}|$.

\subsubsection{Training Data of BLS}


Finally, for every user-item pair $(u,v)$ with known rating in the training rating matrix, the user-item rating collaborative vector $\mathbf{x}_{v}^u$ and the user-item rating $r_{u,v}$ are taken as the input and output of a training sample $(\mathbf{x}_{v}^u,r_{u,v})$ respectively. That is, the training set of the proposed BroadCF method is
\begin{equation}
	\label{eq:D}
	\mathcal{D}=\{(\mathbf{x}_{v}^u, r_{u,v})|\mathbf{R}_{u,v}\neq 0\}.
\end{equation}

\subsection{Broad Learning}

In this subsection, we will describe the mapping function $f$ that maps the rating collaborative
vector  $\mathbf{x}_{v}^u$ into the predicted rating $r_{u,v}$, which will be achieved by BLS. According to~\cite{DBLP:journals/tnn/ChenLS19}, BLS is composed of three modules, namely mapped feature layer, enhanced feature layer, and output layer.

\subsubsection{Mapped Feature Layer}
Given the training set $\mathcal{D}=\{(\mathbf{x}_{v}^u, r_{u,v})|\mathbf{R}_{u,v}\neq 0\}$, the training input matrix of BLS is formed as
\begin{equation}
	\label{eq:X}
	\mathbf{X}=\left[
	\begin{array}{c}
		\mathbf{x}_{v_1}^{u_1} \\
		\mathbf{x}_{v_2}^{u_2} \\
		\vdots \\
		\mathbf{x}_{v_{|\mathcal{D}|}}^{u_{|\mathcal{D}|}} \\
	\end{array}
	\right]\in\mathbb{R}^{|\mathcal{D}|\times (k+l)}
\end{equation}
where $|\mathcal{D}|$ is the training sample number, \textit{i.e.}, there are $|\mathcal{D}|$ user-item pairs $(u_1,v_1), (u_2,v_2), ..., (u_{|\mathcal{D}|},v_{|\mathcal{D}|})$.
The training output vector of BLS is formed as
\begin{equation}
	\label{eq:Y}
	\mathbf{y}=\left[
	\begin{array}{c}
		r_{u_1,v_1} \\
		r_{u_2,v_2} \\
		\vdots \\
		r_{u_{|\mathcal{D}|},v_{|\mathcal{D}|}} \\
	\end{array}
	\right]\in\mathbb{R}^{|\mathcal{D}|\times 1}.
\end{equation}

In the mapped feature layer, the input matrix $\mathbf{X}$ is transformed into the mapped feature matrix $\mathbf{Z}_i\in\mathbb{R}^{|\mathcal{D}|\times d_z}$ as follows
\begin{eqnarray}
	\begin{aligned}
		\mathbf{Z}_i=	
		\phi_i(\mathbf{X}\mathbf{W}_{z_i}+\bm{\beta}_{z_i}), ~~i= 1,2,\cdots,n
	\end{aligned}
\end{eqnarray}
where $d_z$ denotes the dimension of each mapped feature group, $n$ is the number of the mapped feature groups, and $\phi_i$ is the $i$-th nonlinear feature mapping function. In our experiments, $\text{ReLu}$ is used as the nonlinear feature mapping function. In the above procedure, $\mathbf{W}_{z_i}\in\mathbb{R}^{(k+l)\times d_z}$ and $\bm{\beta}_{z_i}\in\mathbb{R}^{|\mathcal{D}|\times d_z}$ are randomly generated matrices {{according to some distributions such as normal distribution}. However, randomness suffers from unpredictability. To overcome the randomness nature, sparse autoencoder can be regarded as an important tool to slightly fine-tune the random features to get a sparse and compact  features. According to~\cite{DBLP:journals/tnn/ChenLS19}, to utilize the properties of the sparse autoencoder, {the alternating direction multiplier method (ADMM) is adopted to learn the tuned weight matrix} by the following iterative steps:}
{
\begin{equation}
\label{eq:sparse w}
\left\{\begin{array}{l}
\hat{\mathbf{W}}_{k+1}=\left(\hat{\mathbf{Z}}_i^{{\top}} \hat{\mathbf{Z}}_i+r \mathbf{I}\right)^{-1}\left(\hat{\mathbf{Z}}_i^{{\top}} \mathbf{X}+r\left(\mathbf{o}_k-\mathbf{u}_k\right)\right) \\
\mathbf{o}_{k+1}=\mathrm{S}_{\lambda / r}\left(\hat{\mathbf{W}}_{k+1}+\mathbf{u}_k\right) \\
\mathbf{u}_{k+1}=\mathbf{u}_k+\left(\hat{\mathbf{W}}_{k+1}-\mathbf{o}_{k+1}\right),
\end{array}\right.
\end{equation}
where $\hat{\mathbf{Z}}_i$ is the desired sparse feature matrix, $\hat{\mathbf{W}}_{k}$ is the sparse autoencoder solution, $\hat{\mathbf{W}}_{0}$, $\mathbf{o}_{0}$ and $\mathbf{u}_{0}$ are {initialized as the zero matrices}, and $\lambda$ is the regularization parameter. }

{In addition, {$r>0$} and {$S_\zeta$} is the soft thresholding operator which is defined as:
\begin{equation}
S_\zeta(a)= \begin{cases}a-\zeta, & a>\zeta \\ 0, & |a| \leq \zeta \\ a+\zeta, & a<-\zeta\end{cases} .
\end{equation}}

{We apply the linear inverse problem in Eq.~(\ref{eq:sparse w}) and fine-tune the initial $\mathbf{W}_{z_i}$ to obtain better features.} In this way, the mapped feature matrices can be obtained, \textit{i.e.},
\begin{equation}
	\label{eq:Z^n}
	\mathbf{Z}^n = [\mathbf{Z}_1|\mathbf{Z}_2\cdots|\mathbf{Z}_n]\in\mathbb{R}^{|\mathcal{D}|\times nd_z}.
\end{equation}

\begin{algorithm}[!t]
	\renewcommand{\algorithmicrequire}{\textbf{Input:}}
	\renewcommand{\algorithmicensure}{\textbf{Output:}}
	\caption{Broad Collaborative Filtering (BroadCF)}
	\label{alg1}
	\textbf{Training part:}
	\begin{algorithmic}[1]
		\STATE \textbf{Input:} $\mathbf{R}$: training rating matrix; $k$: number of the nearest users; $l$: number of the nearest items; $n$: number of mapped feature groups; $d_z$: mapped feature dimension; $m$: number of enhanced feature groups; $d_h$: enhanced feature dimension.
		\STATE Search the $k$ nearest users of each user $u$.
		\STATE Search the $l$ nearest items of each item $v$.
		\FORALL{training user-item pair with $\mathbf{R}_{u,v}\neq 0$}
		\STATE Generate user-item rating collaborative vector $\mathbf{x}^u_{v}$ via Eq.~(\ref{eq:x_{v}^u}).
		\ENDFOR
		\STATE Generate the training set $\mathcal{D}$ via Eq. (\ref{eq:D}).
		\STATE Generate the training input matrix $\mathbf{X}$ and training output vector $\mathbf{y}$ via Eq. (\ref{eq:X}) and Eq. (\ref{eq:Y}) respectively.
		\STATE Generate the mapped feature matrices $\mathbf{Z}^n$ and enhanced feature matrices $\mathbf{H}^m$ via Eq. (\ref{eq:Z^n}) and Eq. (\ref{eq:H^m}) respectively.
		\STATE Calculate the trainable weight vector $\mathbf{w}$ via Eq.~(\ref{eq:W}).
		\STATE \textbf{Output:}  The trainable weight vector $\mathbf{w}$.
	\end{algorithmic}
	\textbf{Testing part:}
	\begin{algorithmic}[1]
		\STATE \textbf{Input:} A user-item pair $(u,v)$ with unknown rating.
		\STATE Generate user-item rating collaborative vector $\mathbf{x}^u_{v}$ via Eq.~(\ref{eq:x_{v}^u}).
		\STATE Feed $\mathbf{x}^u_{v}$ into the broad learning procedure to generate the predicted user-item rating $\hat{r}^u_{v}$.
		\STATE \textbf{Output:} The predicted rating $\hat{r}^u_{v}$.
	\end{algorithmic}
\end{algorithm}

\subsubsection{Enhanced Feature Layer}
After obtaining $\mathbf{Z}^n$, the enhanced feature matrix $\mathbf{H}_i\in\mathbb{R}^{|\mathcal{D}|\times d_h}$ is calculated as follows,
\begin{eqnarray}
	\begin{aligned}
		\mathbf{H}_i= 	
		\xi_i(\mathbf{Z}^n\mathbf{W}_{h_i}+\bm{\beta}_{h_i}), ~~i= 1,2,\cdots,m
	\end{aligned}
\end{eqnarray}
where $d_h$ denotes the dimension of each enhanced feature group, $m$ is the number of the enhanced feature groups, and $\xi_i$ is the $i$-th nonlinear feature enhancement function. In our experiments, $\text{ReLu}$ is used as the nonlinear feature enhancement function. In the above procedure, $\mathbf{W}_{h_i}\in\mathbb{R}^{nd_z\times d_h}$ and $\bm{\beta}_{h_i}\in\mathbb{R}^{|\mathcal{D}|\times d_h}$ are randomly generated matrices.
In this way, the enhanced feature matrices is obtained, \textit{i.e.},
\begin{equation}
	\label{eq:H^m}
	\mathbf{H}^m = [\mathbf{H}_1|\mathbf{H}_2\cdots|\mathbf{H}_m]\in\mathbb{R}^{|\mathcal{D}|\times md_h}.
\end{equation}

\subsubsection{Output Layer}

After obtaining the mapped feature matrices $\mathbf{Z}^n$ and the enhanced feature matrices $\mathbf{H}^m$, the output vector $\mathbf{y}\in\mathbb{R}^{|\mathcal{D}|\times 1}$ can be predicted as
\begin{eqnarray}
	\begin{aligned}
		\begin{aligned}
			\mathbf{y}   &=  \left[\mathbf{Z}^n |\mathbf{H}^m \right]\mathbf{w},
		\end{aligned}
	\end{aligned}
\end{eqnarray}
where $\mathbf{w}\in\mathbb{R}^{(nd_z+md_h)\times 1}$ is the trainable weight vector that maps the concatenated mapped and enhanced features into the output $\mathbf{y}$.

{In the training procedure, {BLS} does not require an iterative training process, and} the only trainable weight vector $\mathbf{w}$ can be solved by applying the ridge regression approximation of pseudoinverse $\left[\mathbf{Z}^n |\mathbf{H}^m \right]^+$, \textit{i.e.},
\begin{equation}
	\label{eq:W}
	\mathbf{w}=\left[\mathbf{Z}^n |\mathbf{H}^m \right]^+\mathbf{y},
\end{equation}
{where
\begin{equation}
\left[\mathbf{Z}^n |\mathbf{H}^m \right]^+=\lim _{\lambda \rightarrow 0}\left(\lambda \mathbf{I}+\left[\mathbf{Z}^n |\mathbf{H}^m \right] \left[\mathbf{Z}^n |\mathbf{H}^m \right]^{\top}\right)^{-1} \left[\mathbf{Z}^n |\mathbf{H}^m \right]^{\top} .
\end{equation}
}

{The training of BLS {is composed of} two main stages: 1) randomly generating the weights of the mapped {feature} layer and the {e}nhanced {feature} layer, and 2) calculating the weights between the hidden layer and the output layer by pseudo-inverse. In other words, the weights of the mapped and enhanced layers of BLS are randomly generated, and only the weight{s} of the link{s} between the final output layer and the hidden layer need to be obtained through training. While BLS does not require an iterative training process, the weight{s} are obtained by ridge regression approximation pseudo-inverse computation.}

{The general idea behind BroadCF is that potential user and item feature matrices are essentially generated by a factori{z}ation-based approach. In contrast to other matrix decomposition models, BroadCF generates potential user feature matrices and item feature matrices through a preprocessing procedure, and the predicted ratings are obtained by using BLS to learn the nonlinear relationship between users and items.}

\subsection{Rating Prediction{,}  Algorithm Summary {and Analysis}}

In the testing procedure, for every user-item pair $(u,v)$ with unknown rating, the predicted rating can be obtained by feeding the user-item rating collaborative vector $\mathbf{x}_{v}^u\in\mathbb{R}^{1\times (k+l)}$ into the broad learning procedure, which outputs the predicted user-item rating $\hat{r}_{u,v}$. For clarity, Algorithm~\ref{alg1} summarizes the main procedure of BroadCF.

{The computational complexity of BroadCF {mainly contains} two parts: the data preprocessing module and the BLS module. The computational complexity of the data preprocessing module is $O((\left| \mathcal{U} \right|)^2+(\left| \mathcal{V} \right|)^2)$. The computational complexity of the BLS module consists of three main parts: generation of feature nodes in the mapped feature layer, generation of feature nodes in the enhanced feature layer, and computation of the pseudoinverse. The mapped feature layer is performed by mapping the input matrix into $n$ mapped feature groups, where each mapped feature has dimension $d_z$. The computational complexity of this step is $O(\mathcal{D}(k+l)nd_z)$. The enhanced feature layer maps the mapped feature matrix to the enhanced feature matrix, and the computational complexity of this step is $O(\mathcal{D}nmd_zd_h)$. The pseudoinverse computation is performed by solving the output weight matrix obtained by ridge regression parsing, and the computational complexity of this step is $O(\mathcal{D}(nm+d_zd_h)^2$. Therefore, the computational complexity of BroadCF is computed as $O((\left| \mathcal{U} \right|)^2+(\left| \mathcal{V} \right|)^2+\mathcal{D}(k+l)nd_z+\mathcal{D}nmd_zd_h+\mathcal{D}(nm+d_zd_h)^2)$.}

{The only trainable parameters to be stored in BroadCF are the weight vector $\mathbf{w}\in\mathbb{R}^{(nd_z+md_h)\times 1}$, which is independent of the input data size and dimension. Therefore, {the number of trainable parameters to be stored during the training procedure of BroadCF is $(nd_z+md_h)$}.}

\begin{table}[!t]
	\caption{Statistic information of the seven datasets.}
	\label{tab:dataset}
	\vskip -0.1in
	\centering
	\resizebox{1\linewidth}{!}{
		\begin{tabular}{crrrrr}
			\hline
			\multicolumn{1}{c}{\textbf{Datasets}} & \multicolumn{1}{c}{\textbf{\#Users}} & \multicolumn{1}{c}{\textbf{\#Items}} & \multicolumn{1}{c}{\textbf{\#Ratings}} & \multicolumn{1}{c}{\textbf{Sparsity}} & \multicolumn{1}{c}{\textbf{Scale}} \\
			\hline
			ml-la & 610   & 9,724  & 100,836  & 98.30\% & [1, 5] \\
			Amazon-DM & 1000 & 4,796 & 11,677 & 99.90\% & [1, 5] \\
			Amazon-GGF & 2,500 & 7,527   & 11,532 & 99.96\% & [1, 5] \\
			Amazon-PLG & 5,000 & 5,400 & 12,429 & 99.98\% & [1, 5] \\		
			Amazon-Automotive & 5,000 & 6,596   & 12,665 & 99.98\% & [1, 5] \\
			Amazon-Baby & 6,000 & 5,428   & 17,532 & 99.98\% & [1, 5] \\
			Amazon-IV & 429,622 & 23,966 & 583,933 & 99.98\% & [1, 5] \\
			\hline
		\end{tabular}
	}
	\vskip -0.1in
\end{table}%

\begin{table*}[!t]
	\caption{The comparison results on {the rating} prediction performance: The mean value and standard deviation (std. dev.) of RMSE achieved by running each algorithm five times on each dataset.}
	\label{tab:rmse}
	\vskip -0.1in
	\centering
	\resizebox{1\linewidth}{!}{
		\begin{tabular}{ccccccccccccccccc}
			\hline
			\multicolumn{1}{c}{\textbf{Methods}} & \multicolumn{2}{c}{\textbf{ml-la}} & \multicolumn{2}{c}{\textbf{Amazon-DM}} & \multicolumn{2}{c}{\textbf{Amazon-GGF}}  & \multicolumn{2}{c}{\textbf{Amazon-PLG}} & \multicolumn{2}{c}{\textbf{Amazon-Automotive}} & \multicolumn{2}{c}{\textbf{{Amazon-Baby}}} & \multicolumn{2}{c}{\textbf{{Amazon-IV}}} \\
			\hline
&Mean&Std. dev.&Mean&Std. dev.&Mean&Std. dev.&Mean&Std. dev.&Mean&Std. dev.&Mean&Std. dev.&Mean &Std. dev.\\
\hline
			\textbf{PMF} & 1.0016 & 0.0029 & 2.2170 & 0.0122  & 3.2144 & 0.0263 &  3.0658 & 0.0346 &  3.2621 & 0.0329 & 2.6281 & 0.0225 & NA  & NA  \\
			\textbf{NeuMF} & 2.3643 & 0.0042 & 2.0587 & 0.0253 & 2.3508 & 0.0497 & 1.7665 & 0.0391 & 2.9946 & 0.0310 & 2.9796 & 0.0374 & 1.2749  & 0.0258 \\
			\textbf{DMF} & 1.6325 & 0.0034 & 3.6161 & 0.0215 & 4.1642 & 0.0328 & 4.1030 & 0.0216 & 4.1488 & 0.1088 & 3.2546 & 0.0154 & NA & NA  \\
			\textbf{DeepCF} & 2.1133 & 0.0094 & 1.7025 & 0.0217 & 1.2487 & 0.0314 & 1.6426 & 0.0452 & 1.6352 & 0.0449 & 1.6669 & 0.0441 & NA  & NA  \\
			\textbf{BPAM} & 0.8484 & 0.0035 & 1.2509 & 0.0132 & 1.7281 & 0.0274 & 1.9805 & 0.0328  & 1.9805 & 0.0373 & 0.8678 & 0.0301 & 2.4521   & 0.0324  \\
			\textbf{SHT} & 1.5023 & 0.0096 & 2.2758 & 0.1082 & 1.9724 & 0.0501& 1.6721 & 0.1120 & 1.4503 & 0.0731 & 2.0094 & 0.0636 & NA  & NA\\
   			\textbf{NGCF} & 1.0441 & 0.0196 & 2.7570 & 0.0447 & 2.3956 & 0.0341 & 3.6154 & 0.0386 & 2.1711 & 0.0186 & 2.3834 & 0.0275 &  NA & NA    \\
   			\textbf{LigthGCN} & 0.9221 & 0.0630 & 1.3119 & 0.0504 & 2.4928 & 0.0790 & 2.0558 & 0.0862  & 2.1790 & 0.0735 & 2.7775 &  0.0568 &  NA & NA \\
			\textbf{BroadCF} & 0.7220 & 0.0363 & 0.7292 & 0.1542 & 0.9683 & 0.1110 & 0.9877 & 0.0698 & 0.8570 & 0.0155 & 0.8528 & 0.0426 & 0.7933  & 0.0542 \\
			\hline
	\end{tabular}}
 \vskip -0.1in
\end{table*}

\begin{table*}[!t]
	\caption{The comparison results on {the rating} prediction performance: The mean value and standard deviation (std. dev.) of MAE achieved by running each algorithm five times on each dataset.}
	\label{tab:mae}
	\vskip -0.1in
	\centering
	\resizebox{1\linewidth}{!}{
		\begin{tabular}{ccccccccccccccccc}
			\hline
			\multicolumn{1}{c}{\textbf{Methods}} & \multicolumn{2}{c}{\textbf{ml-la}} & \multicolumn{2}{c}{\textbf{Amazon-DM}} & \multicolumn{2}{c}{\textbf{Amazon-GGF}}  & \multicolumn{2}{c}{\textbf{Amazon-PLG}} & \multicolumn{2}{c}{\textbf{Amazon-Automotive}} & \multicolumn{2}{c}{\textbf{{Amazon-Baby}}} & \multicolumn{2}{c}{\textbf{{Amazon-IV}}} \\
			\hline
&Mean&Std. dev.&Mean&Std. dev.&Mean&Std. dev.&Mean&Std. dev.&Mean&Std. dev.&Mean&Std. dev.&Mean&Std. dev.\\
			\hline
			\textbf{PMF} & 0.7525 & 0.0026 & 1.6670 & 0.0107 & 2.9038 & 0.0320 & 2.6038 & 0.0281 & 2.8330 & 0.0269 & 2.1097 & 0.0117 & NA  & NA  \\
			\textbf{NeuMF} & 1.9791 & 0.0032 & 1.3896 & 0.0221 & 1.6070 & 0.0374 & 1.7446 & 0.0323 & 2.5148 & 0.0215 & 2.4489 & 0.0308 & 1.0138  & 0.0211  \\
			\textbf{DMF} & 1.2297 & 0.0037 & 3.4477 & 0.0275 & 3.9433 & 0.0305 & 3.8417 & 0.0226 & 3.9055 & 0.1004 & 3.2546 & 0.0131 & NA  & NA  \\
			\textbf{DeepCF} & 1.6345 & 0.0082 & 1.6202 & 0.0188 & 1.2237 &  0.0296 & 1.6189 & 0.0360 & 1.6798 &  0.0391 & 1.5738 & 0.0319 & NA  & NA  \\
			\textbf{BPAM} & 0.6570 & 0.0024& 1.5800 & 0.0175 & 1.2425 & 0.0212& 1.5440 & 0.0290 & 0.6790 & 0.0326 & 1.4900 & 0.0257 & 1.9529  & 0.0289  \\
			\textbf{SHT} & 1.4391 & 0.0105 & 1.7279 & 0.0658 & 1.5720 & 0.0419& 1.4839 & 0.0914 & 1.2942 & 0.0739 & 1.4900 & 0.0541 & NA  & NA  \\
      		\textbf{NGCF} & 0.7318 & 0.0347 & 1.5203 & 0.0101 & 1.1478 & 0.0371 & 2.6143 & 0.0441 & 0.9428 & 0.0868 & 1.1361 & 0.0194 &  NA & NA \\
   			\textbf{LigthGCN} & 0.7282 & 0.0169  & 1.1828 & 0.0226 & 1.5161 & 0.0563 & 1.9520 &  0.0357 & 1.6649 & 0.0493 & 1.4371 & 0.0481 &  NA & NA \\
			\textbf{BroadCF} & 0.5353 & 0.0274 & 0.5151 & 0.1320 & 0.5903 & 0.0563 & 0.6512 & 0.0698 & 0.5202 & 0.0585 & 0.5672 & 0.0432 & 0.3994  & 0.0507  \\
			\hline
	\end{tabular}}
 \vskip -0.1in
\end{table*}

\begin{table*}[!t]
	\caption{{The comparison results on {the recommendation} performance: The mean value {and standard deviation (std. dev.)} of {NDCG@10} achieved by running each algorithm five times on each dataset.} }
	\label{tab:NDCG@10}
	\vskip -0.1in
	\centering
	\resizebox{1\linewidth}{!}{
		\begin{tabular}{ccccccccccccccccc}
			\hline
			\multicolumn{1}{c}{\textbf{Methods}} & \multicolumn{2}{c}{\textbf{ml-la}} & \multicolumn{2}{c}{\textbf{Amazon-DM}} & \multicolumn{2}{c}{\textbf{Amazon-GGF}}  & \multicolumn{2}{c}{\textbf{Amazon-PLG}} & \multicolumn{2}{c}{\textbf{Amazon-Automotive}} & \multicolumn{2}{c}{\textbf{{Amazon-Baby}}} & \multicolumn{2}{c}{\textbf{{Amazon-IV}}} \\
			\hline
&Mean&Std. dev.&Mean&Std. dev.&Mean&Std. dev.&Mean&Std. dev.&Mean&Std. dev.&Mean&Std. dev.&Mean&Std. dev.\\
			\hline
			\textbf{PMF} & 0.3162  & 0.0012 & 0.3623 & 0.0021 & 0.1943 & 0.0015 & 0.1173 & 0.0009 & 0.1067 & 0.0011 & 0.2294 & 0.0013 & NA & NA  \\
			\textbf{NeuMF} & 0.2435 & 0.0006 & 0.2607 & 0.0008 & 0.3824 & 0.0014 & 0.3085 & 0.0019 & 0.2970 & 0.0017 & 0.2602 & 0.0017 & 0.2893 & 0.0021  \\
			\textbf{DMF} & 0.4470 & 0.0106 & 0.3646 & 0.0043 & 0.2959 & 0.0028 & 0.2592 & 0.0016 & 0.2924 & 0.0011 &  0.3078 & 0.0052 & NA  & NA\\
			\textbf{DeepCF} & 0.3743 & 0.0042 & 0.3856 & 0.0121 & 0.3133 & 0.0043 & 0.3120 & 0.0029 & 0.2921 & 0.0031 & 0.1210 & 0.0012 & NA  & NA  \\
			\textbf{BPAM} & 0.2139 & 0.0072 & 0.2610 & 0.0062 & 0.3019 & 0.0025 & 0.2597 & 0.0092 & 0.2233 & 0.0201 & 0.2821 & 0.0351 & 0.2410  & 0.0108  \\
			\textbf{SHT} & 0.2045  & 0.0281 & 0.2670  & 0.0042 & 0.2842 & 0.0054  & 0.2277 & 0.0028  & 0.2017 & 0.0036  & 0.2250 & 0.0041 & NA  & NA \\
                \textbf{NGCF} & 0.3873 & 0.0219 & 0.3468 & 0.0341 & 0.3218 & 0.0214 & 0.3034 & 0.0733 & 0.2944 & 0.0536 & 0.2883 & 0.0825 &  NA & NA \\
   			\textbf{LigthGCN} & 0.4429 & 0.0322 & 0.4080 & 0.0493 & 0.3739 & 0.0152 & 0.3068 & 0.0274 & 0.2942 & 0.0133 & 0.2826 & 0.0667 & NA  & NA \\
			\textbf{BroadCF} & 0.4765 & 0.0544 & 0.3716 & 0.0703 & 0.3484 & 0.0289 & 0.3142 & 0.0353 & 0.2767 & 0.0756 & 0.2951 & 0.0589 & 0.2699 & 0.0138 \\
			\hline
 \end{tabular}}
 \vskip -0.1in
\end{table*}

{
\begin{table*}[!t]
	\caption{The comparison results on {the recommendation} performance: The mean value {and standard deviation (std. dev.)} of HR@10 achieved by running each algorithm five times on each dataset.}
	\label{tab:HR@10}
	\vskip -0.1in
	\centering
	\resizebox{1\linewidth}{!}{
		\begin{tabular}{ccccccccccccccccc}
			\hline
			\multicolumn{1}{c}{\textbf{Methods}} & \multicolumn{2}{c}{\textbf{ml-la}} & \multicolumn{2}{c}{\textbf{Amazon-DM}} & \multicolumn{2}{c}{\textbf{Amazon-GGF}}  & \multicolumn{2}{c}{\textbf{Amazon-PLG}} & \multicolumn{2}{c}{\textbf{Amazon-Automotive}} & \multicolumn{2}{c}{\textbf{{Amazon-Baby}}} & \multicolumn{2}{c}{\textbf{{Amazon-IV}}} \\
			\hline
&Mean&Std. dev.&Mean&Std. dev.&Mean&Std. dev.&Mean&Std. dev.&Mean&Std. dev.&Mean&Std. dev.&Mean&Std. dev.\\
			\hline
			\textbf{PMF} & 0.6737 & 0.0073  & 0.8098  & 0.0125  & 0.4785 & 0.0047  & 0.2364 & 0.0035  & 0.2122 & 0.0011   & 0.6206 & 0.0082 & NA  & NA \\
			\textbf{NeuMF} & 0.3983 & 0.0179  & 0.4444 & 0.0093  & 0.5518 & 0.0131  & 0.5375 & 0.0196  & 0.5537 & 0.0371 & 0.4404 & 0.0072 & 0.5607 & 0.0270   \\
			\textbf{DMF} & 0.8295 & 0.0226 & 0.6786 & 0.0118 & 0.4789 & 0.0139 & 0.3906 & 0.0126 & 0.4568 & 0.0143 & 0.5898 & 0.0187 & NA  & NA  \\
			\textbf{DeepCF} & 0.4934 & 0.0076 & 0.7517 & 0.0137 & 0.4789 & 0.0146 & 0.5979 & 0.0164 & 0.7889 & 0.0427 & 0.2295 & 0.0104  & NA  & NA \\
			\textbf{BPAM} & 0.3513 & 0.0063  & 0.3403 & 0.0047   & 0.3105 & 0.0140 & 0.4430 & 0.0119 & 0.4152 & 0.0230 & 0.3842 & 0.0808  & 0.3462  & 0.0262  \\
			\textbf{SHT} & 0.2046  & 0.0320  & 0.3772 & 0.0337  & 0.5019 & 0.0133   & 0.4061  & 0.0250  & 0.4286  & 0.0152 & 0.3386  & 0.0053 & NA  & NA  \\
      		\textbf{NGCF} & 0.7393 & 0.0228 & 0.6373 & 0.0183 & 0.6139 & 0.0492 & 0.6033 & 0.0314 & 0.7551 & 0.0835 & 0.5516 & 0.0344 &  NA &  NA\\
   			\textbf{LigthGCN} & 0.7774 & 0.0462 & 0.8123 & 0.0379 & 0.6017 & 0.0169 & 0.5898 & 0.0327 & 0.7556 & 0.0193 & 0.6022 & 0.0540 &  NA & NA \\
			\textbf{BroadCF} & 0.8229 & 0.0352 & 0.7927 & 0.0517 & 0.7599 & 0.0468 & 0.6541 & 0.0910 & 0.7927 & 0.0270 & 0.5792 & 0.0994 & 0.6439 & 0.0114 \\
			\hline
	\end{tabular}}
 \vskip -0.1in
\end{table*}}

\section{Experiments}
\label{sec:Experiments}

In this section, extensive experiments will be conducted to confirm the effectiveness of the proposed BroadCF method. All experiments are conducted in Python and run on a server with an Intel Core i9-10900 CPU, GeForce RTX 3090, and 256GB of RAM.

\subsection{Experimental Settings}

\subsubsection{Datasets}

In our experiments, seven real-world publicly available datasets are used, which are obtained from the two main sources.
\begin{enumerate}
	\item
	MovieLens\footnote{\url{https://grouplens.org/datasets/movielens/}}: The MovieLens dataset contains rating data of movies by users. For experimental purpose, we use the ml-latest dataset (\textit{abbr.} ml-la). {The Movie{L}ens dataset contains only ratings, which {are} used in our experiments.}
	\item
	Amazon\footnote{\url{http://jmcauley.ucsd.edu/data/amazon/}}: It is an upgraded version of the Amazon review dataset for 2018~\cite{DBLP:conf/www/HeM16}, which contains reviews (ratings, text, and helpful polls), product meta-data (category information, brand and image features, price, and descriptions) and links (also view/buy charts). {Although the Amazon dataset has different interactions, in our experiments we only {use} ratings.} For experimental purpose, we use six datasets, namely Digital-Music (Amazon-DM), Grocery-and-Gourmet-Food (Amazon-GGF), Patio-Lawn-and-Garden (Amazon-PLG), Automotive (Amazon-Automotive), Baby (Amazon-Baby), and Instant-Video (Amazon-IV).
\end{enumerate}
Table~\ref{tab:dataset} summarizes the {statistical} information of the above seven datasets. From the table, we can see that Amazon-IV is relatively very large. On each dataset, the rating information of each user is randomly split into the training set (75\%) and the testing set (25\%). Furthermore, in the training set, a quarter of the ratings are randomly selected as the validation set to tune the hyper-parameters for all the methods.


\subsubsection{Evaluation Measures}


{To evaluate the performance of rating prediction and recommendation, four evaluation measures are used, namely Root Mean Square Error (\textit{abbr.} RMSE), Mean Absolute Error (\textit{abbr.} MAE),  Normalized Discounted Cumulative Gain (\textit{abbr.} NDCG) and Hit Ratio (\textit{abbr.} HR). The first two measures are calculated via computing the rating prediction error \textit{w.r.t.} the ground-truth rating, where smaller values indicate better rating prediction results. And the last two measures are calculated on the top-$K$ recommendations, where larger values indicate better recommendation results. In our experiments, considering the relatively sparse user-item interaction, $K$ is set to {10, i.e., NDCG@10 and HR@10} are reported. In addition, the running cost, including training time, testing time, and trainable parameter storage cost is also used as evaluation criteria.}


\subsection{Comparison Results}
\subsubsection{Baselines}

The proposed BroadCF approach is compared with the following {eight} approaches.
\begin{itemize}
	\item \textbf{PMF}\footnote{\url{https://github.com/xuChenSJTU/PMF}}~\cite{mnih2008probabilistic} is a classical CF approach that considers
	only latent factors and uses matrix decomposition to capture the linear relationships in the user-item pairs.
	\item \textbf{NeuMF}\footnote{\url{https://github.com/ZJ-Tronker/Neural_Collaborative_Filtering-1}}~\cite{He:2017:NCF:3038912.3052569} takes as input the connection of user/item embeddings and uses the MLP model for prediction.
	\item \textbf{DMF}\footnote{\url{https://github.com/RuidongZ/Deep_Matrix_Factorization_Models}}~\cite{xue2017deep} uses a dual-path neural network instead of the linear embedding operation used in the conventional matrix decomposition.
	\item {\textbf{DeepCF}\footnote{\url{https://github.com/familyld/DeepCF}}~\cite{DeepCF} integrates a unified deep collaborative filtering framework with representation learning and matching function learning.}
	\item {\textbf{BPAM}\footnote{\url{https://github.com/xiwd/BPAM}}~\cite{DBLP:journals/tkde/WangXHZHL22}   is a neighborhood-based CF recommendation framework that captures the global influence of a target user’s nearest users on their nearest target set of users by introducing a global attention matrix.}
    \item {\textbf{SHT}\footnote{\url{https://github.com/akaxlh/SHT}}~\cite{DBLP:conf/kdd/Xia0Z22} introduces the hypergraph transformer network for collaborative filtering recommendation via self-supervised learning.}
    {\item {\textbf{NGCF}\footnote{\url{https://github.com/huangtinglin/NGCF-PyTorch}}~\cite{DBLP:conf/sigir/Wang0WFC19} is a collaborative filtering recommendation method based on graph neural networks. It utilizes graph neural networks to extract higher-order interaction information between users and items to improve
    {the} recommendation performance.}
    }
    {\item {\textbf{LightGCN}\footnote{\url{https://github.com/gusye1234/LightGCN-PyTorch}}~\cite{DBLP:conf/sigir/0001DWLZ020} is a lightweight graph convolutional collaborative filtering recommendation method. It uses only the neighbor aggregation part of GCN and weights the higher-order embeddings of users and items learned on all layers to get the predicted rating.}}
\end{itemize}

The hyper-parameters of the above six methods are tuned as suggested by the original authors. In PMF, the factor number is set to 30 for both users and items, and the regularization hyper-parameters are set to 0.001 and 0.0001 for users and items respectively. In NeuMF, the number of predictors for the generalized matrix decomposition part and the embedding sizes for users and items are set as 8 and 16, respectively. In DMF, the predictor number is set to 64 for both users and items. In DeepCF, the sizes of the MLP layers and the DMF layers in the representation learning and the matching function learning are both set to [512, 256, 128, 64]. In BPAM, the numbers of the nearest users and items are tuned in [3, 5, 7, 9]. In SHT, the number of GNN layers and the learning rate are set to 2 and 0.001 respectively. For the deep
models, \textit{i.e.}, NeuMF, DMF, DeepCF{,} {SHT, NGCF and LightGCN}, the batch size is set as 256.  In BroadCF, we set the number of both nearest users and nearest items as 5, \textit{i.e.}, $k=l=5$, the number of both mapped feature groups and enhanced feature groups as 7, \textit{i.e.}, $n=m=7$, and the mapped feature dimension and the enhanced feature dimension as 3, \textit{i.e.}, $d_z=d_h=3$. For the DNNs-based methods, including NeuMF, DMF, DeepCF, {SHT, NGCF and LightGCN}, GPUs are used, while for the rest methods, including PMF, BPAM and BroadCF, GPU is not used. The code of BroadCF is made publicly available at \url{https://github.com/BroadRS/BroadCF}.

\begin{table*}[!t]
	\caption{The comparison results on running cost: The training time (in seconds) consumed by each algorithm on each dataset.}
	\label{tab:training time}
	\vskip -0.1in
	\centering
	\resizebox{1\linewidth}{!}{
		\begin{tabular}{crrrrrrrr}
			\hline
			\multicolumn{1}{c}{\textbf{Methods}} & \multicolumn{1}{c}{\textbf{ml-la}} & \multicolumn{1}{c}{\textbf{Amazon-DM}} & \multicolumn{1}{c}{\textbf{Amazon-GGF}}  & \multicolumn{1}{c}{\textbf{Amazon-PLG}} & \multicolumn{1}{c}{\textbf{Amazon-Automotive}} & \multicolumn{1}{c}{\textbf{{Amazon-Baby}}} & \multicolumn{2}{c}{\textbf{{Amazon-IV}}} \\
			\hline
			\textbf{PMF} & 98.24 & 202.92 & 1132.87 & 4072.20 & 4311.29 & 1100.17 & NA \\
			\textbf{NeuMF} & 6030.51 & 120.35 & 96.77 & 94.89 & 111.31 & 146.69 &  62129.98  \\
			\textbf{DMF} & 2478.40 & 71.63 & 104.14 & 148.44 & 132.85 & 250.35 & NA  \\
			\textbf{DeepCF} & 2071.06 & 210.10 & 322.90 & 389.22 & 392.46 & 539.62 & NA  \\
			\textbf{BPAM} & 996.48 & 215.99 & 231.66 & 210.94 & 221.24 & 261.61 & 465016.20  \\
			\textbf{SHT} & 59.13 & 32.14 & 61.79 & 121.48 & 144.73 & 148.34 & NA \\
      		\textbf{NGCF} & 293.74 & 54.03 & 107.58 & 138.88 &  149.92 & 146.15  & NA \\
   			\textbf{LigthGCN} & 237.82 & 31.21 & 91.88 &  113.10   & 102.24 & 123.77  & NA \\
			\textbf{BroadCF} & 48.72 & 7.12 & 22.23 & 69.34 & 88.13 & 94.62 & 311054.57 \\
			\hline
	\end{tabular}}
 \vskip -0.1in
\end{table*}

\begin{table*}[!t]
	\caption{The comparison results on running cost: The testing time (in seconds) consumed by each algorithm on each dataset.}
	\label{tab:testing time}
	\vskip -0.1in
	\centering
	\resizebox{1\linewidth}{!}{
		\begin{tabular}{crrrrrrrr}
			\hline
			\multicolumn{1}{c}{\textbf{Methods}} & \multicolumn{1}{c}{\textbf{ml-la}} & \multicolumn{1}{c}{\textbf{Amazon-DM}} & \multicolumn{1}{c}{\textbf{Amazon-GGF}}  & \multicolumn{1}{c}{\textbf{Amazon-PLG}} & \multicolumn{1}{c}{\textbf{Amazon-Automotive}} & \multicolumn{1}{c}{\textbf{{Amazon-Baby}}} & \multicolumn{2}{c}{\textbf{{Amazon-IV}}}\\
			\hline
			\textbf{PMF} & 0.03 & 0.01 & 0.01 & 0.01 & 0.01 & 0.01 & NA \\
			\textbf{NeuMF} & 301.40 & 118.53 & 189.72 & 331.24 & 464.19 & 450.14 &  10825.30 \\
			\textbf{DMF} & 21.11 & 7.95 & 29.69 & 95.61 & 73.82 & 126.90 & NA \\
			\textbf{DeepCF} & 427.09 & 67.44 & 102.65 & 120.89 & 94.47 & 171.50 & NA \\
			\textbf{BPAM} & 279.80 & 29.20 & 29.58 & 28.62 & 30.51 & 41.55 &  151207.13 \\
			\textbf{SHT} & 5.84 & 2.12 & 6.72 & 20.17 & 27.12 & 26.12 & NA \\
      		\textbf{NGCF} & 6.53 & 4.28 & 10.76  & 25.58 & 26.56 &  31.28 & NA \\
   			\textbf{LigthGCN} & 5.52 & 3.30 & 9.66 &  22.97  & 23.29 &  29.41 & NA \\
			\textbf{BroadCF} & 3.86 & 1.21 & 5.01 & 16.59 & 20.90 & 22.08 & 77766.55 \\
			\hline
	\end{tabular}}
 \vskip -0.1in
\end{table*}

\begin{table*}[!t]
	\caption{The comparison results on running cost: The number of trainable parameters to be stored during the training procedure of each algorithm on each dataset.}
	\label{tab:storage costs}
	\vskip -0.1in
	\centering
	\resizebox{1\linewidth}{!}{
		\begin{tabular}{crrrrrrrr}
			\hline
			\multicolumn{1}{c}{\textbf{}} & \multicolumn{1}{c}{\textbf{ml-la}} & \multicolumn{1}{c}{\textbf{Amazon-DM}} & \multicolumn{1}{c}{\textbf{Amazon-GGF}}  & \multicolumn{1}{c}{\textbf{Amazon-PLG}} & \multicolumn{1}{c}{\textbf{Amazon-Automotive}} & \multicolumn{1}{c}{\textbf{{Amazon-Baby}}} & \multicolumn{2}{c}{\textbf{{Amazon-IV}}} \\
			\hline
			\textbf{PMF} & 310,080 & 173,910 & 300,840 & 312,030 & 347,910 & 342,900 & NA \\
			\textbf{NeuMF} & 392,601 & 234,521 & 403,761 & 418,681 & 466,521 & 459,801 &  18,038,161 \\
			\textbf{DMF} & 12,327,168 & 33,519,232 & 40,030,464 & 48,012,672 & 20,921,728 & 29,942,272 & NA\\
			\textbf{DeepCF} & 39,266,060 & 97,444,766 & 115,947,300 & 137,582,762 & 62,738,578 & 86,977,048 & NA \\
			\textbf{BPAM} & 43,615 & 31,500 & 78,750 & 157,500 & 157,500 & 189,000 & 7,518,367 \\
			\textbf{SHT} & 509,025 & 363,777 & 511,105 & 499,169 & 549,377 & 544,001 & NA \\
      		\textbf{NGCF} & 688,257 & 397,761 & 668,545 & 692,417 & 768,961  & 758,209 &  NA  \\
   			\textbf{LigthGCN} & 661,376 & 370,880 & 641,664 &  665,536 &  742,080  & 731,328  & NA \\
			\textbf{BroadCF} & 3,125 & 3,125 & 3,125 & 3,125 & 3,125 & 3,125 & 3,125 \\
			\hline
	\end{tabular}}
 \vskip -0.1in
\end{table*}

\subsubsection{Comparison on {Rating Prediction and Recommendation}}
The comparison results on the prediction performance measured by RMSE and MAE are listed in Table~\ref{tab:rmse} and Table~\ref{tab:mae} respectively. {Likewise, the comparison results in terms of NDCG@10 and HR@10 are listed in Table~\ref{tab:NDCG@10} and Table~\ref{tab:HR@10}.} In the comparison experiment, each method is run 5 times on each dataset, and the mean value and standard deviation (std. dev.) over the 5 runs are listed. Out-of-memory errors occur when training PMF, DMF, DeepCF{,} {SHT, NGCF and LightGCN} on the largest Amazon-IV dataset. This is because the four models take the user-item rating matrix as input, but for the Amazon-IV dataset, the entire rating matrix takes up to $32\cdot|\mathcal{U}| \cdot |\mathcal{V}|$ bits of memory, which is about 308.64 GB, exceeding 256 GB memory of the experimental machine. As a result, no result is reported in the corresponding entries of the comparison result tables (\textit{i.e.} NA). From the two tables, we have the following observations.

Overall, the values of RMSE and MAE achieved by the proposed BroadCF algorithm are much smaller than those by the other algorithms, including the classical CF method (PMF), four DNNs-based algorithms (DMF, NeuMF, DeepCF, {SHT, NGCF and LightGCN}), and one BP neural network-based method (BPAM). {Also, the NDCG@10 and HR@10 achieved by the proposed BroadCF algorithm {are the largest} on most of the datasets.} In particular, compared with PMF, the proposed BroadCF algorithm achieves at least 27.91\% and 28.86\% improvements of RMSE and MAE, which is a relatively significant performance improvement. The main reason is that PMF only captures the linear latent factors, \textit{i.e.} the linear relationships in the user-item pairs, while BroadCF is able to capture the nonlinear relationships in the user-item pairs. Compared with the {six} DNNs-based algorithms, BroadCF also obtains significant improvements of {RMSE, MAE, NDCG@10 and HR@10}, which is a more significant performance improvement. The main reason may be that, although both of the DNNs-based algorithms and BroadCF are able to capture the nonlinear relationships in the user-item pairs, the DNNs-based algorithms easily encounter the overfitting issue, \textit{i.e.}, given the very large number of trainable parameters, the testing performance cannot be improved compared with the training performance. As a contrast, compared with BPAM, which is a lightweight neural network based method, the prediction performance of BroadCF is still better than that of BPAM, although the {RMSE and MAE} improvement is not such significant. {However, the improvement of BroadCF over BPAM in NDCG@10 and HR@10 is significant.} As will be shown in the comparison on running cost, the number of trainable parameters in BPAM is relatively small compared with those of the DNNs-based algorithms, although still larger than that of BroadCF. Overall, the comparison results on the prediction performance have demonstrated the effectiveness of the proposed BroadCF method.

	\begin{figure*}[!t]
		\centering
		\subfigure[{ml-la}]{
			\centering
			\includegraphics[width=0.24\linewidth]{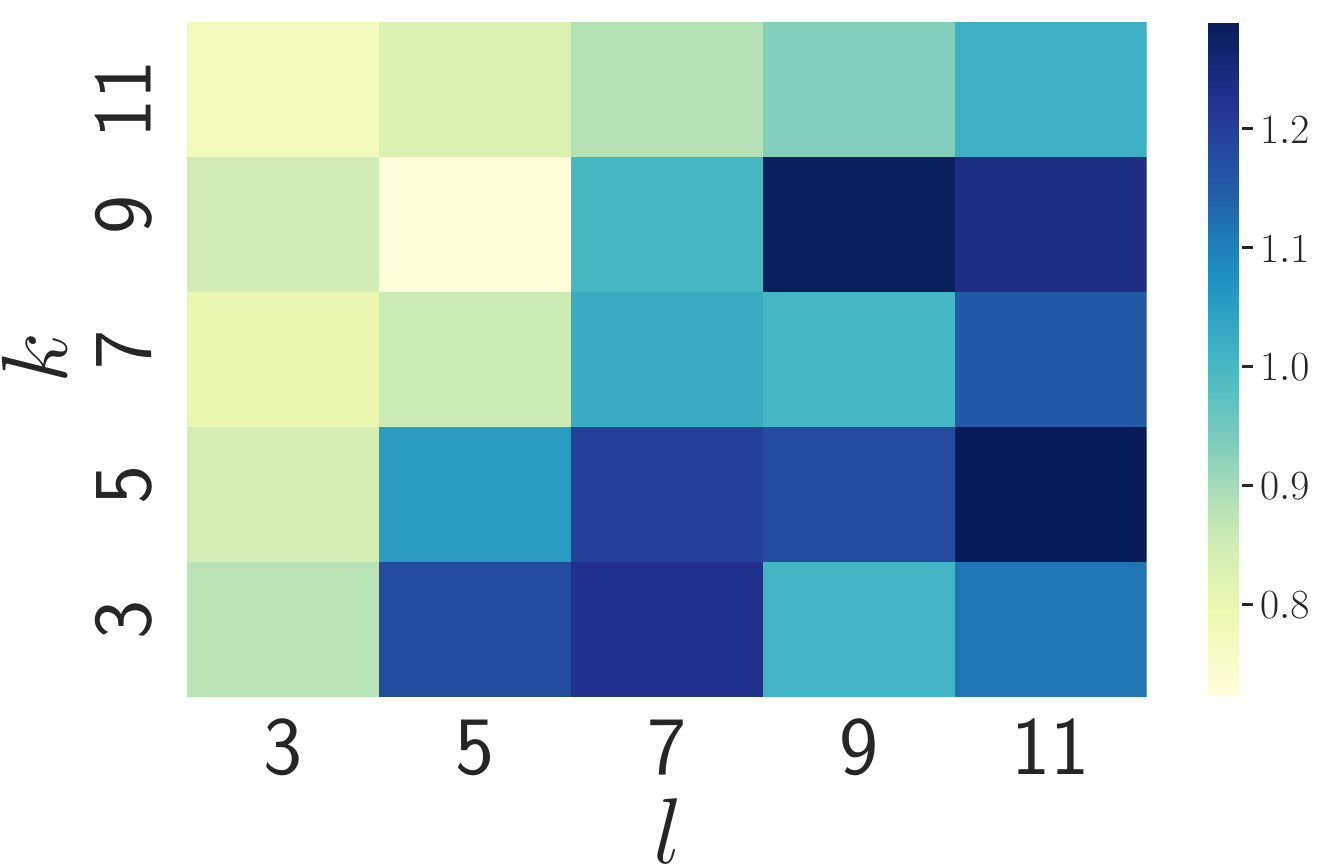}
		}%
		\subfigure[{Amazon-DM}]{
			\centering
			\includegraphics[width=0.24\linewidth]{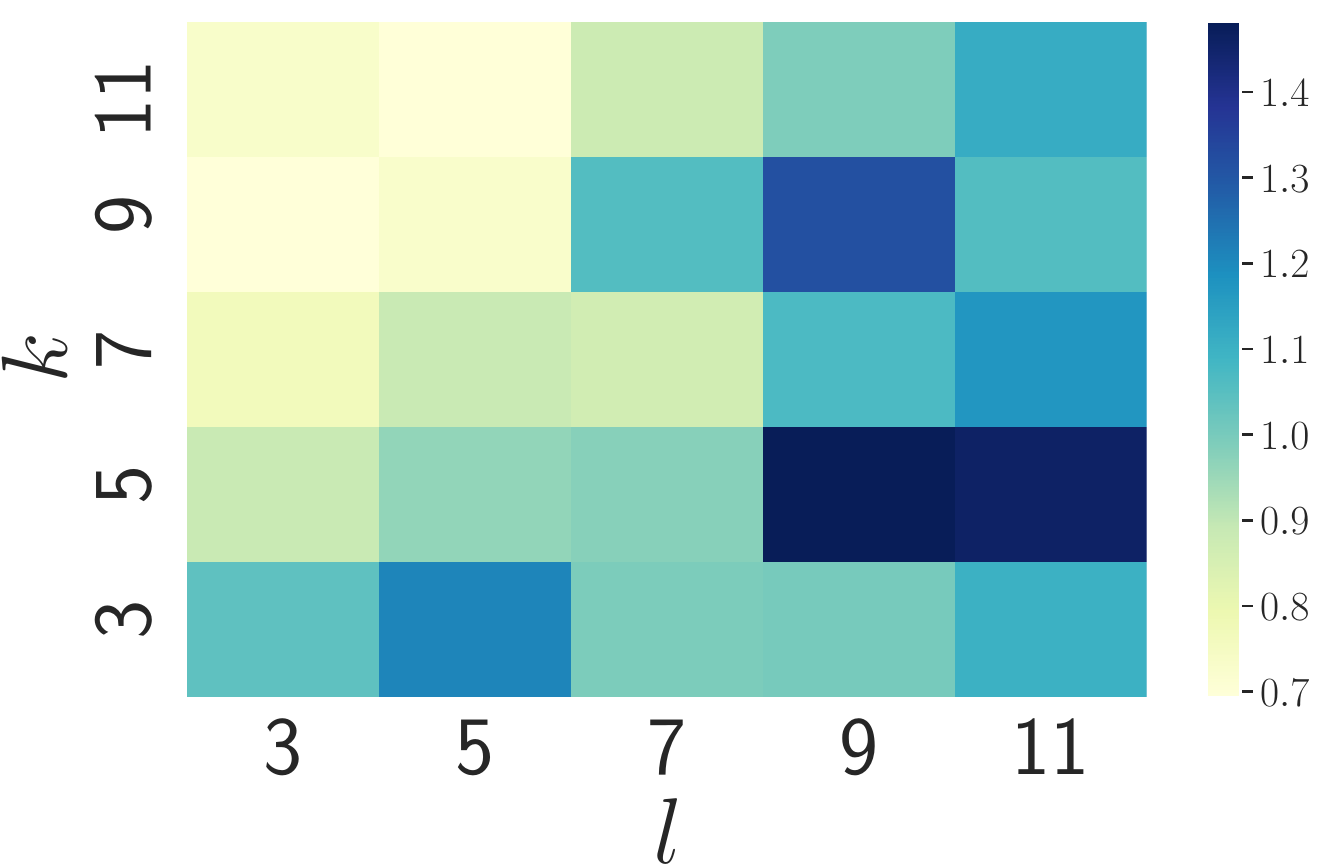}
		}%
		\subfigure[{Amazon-GGF}]{
			\centering
			\includegraphics[width=0.24\linewidth]{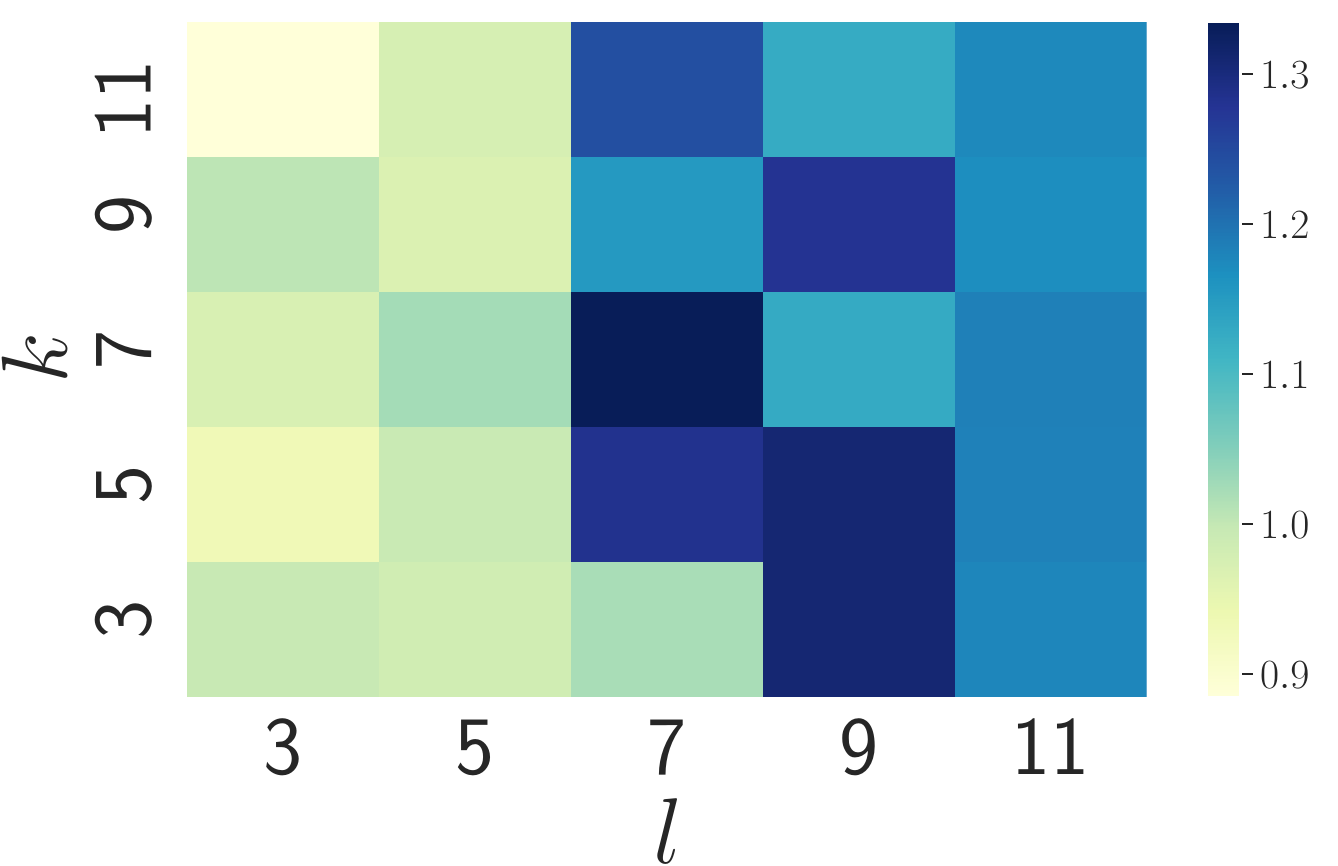}
		}%
		\subfigure[{Amazon-PLG}]{
			\centering
			\includegraphics[width=0.24\linewidth]{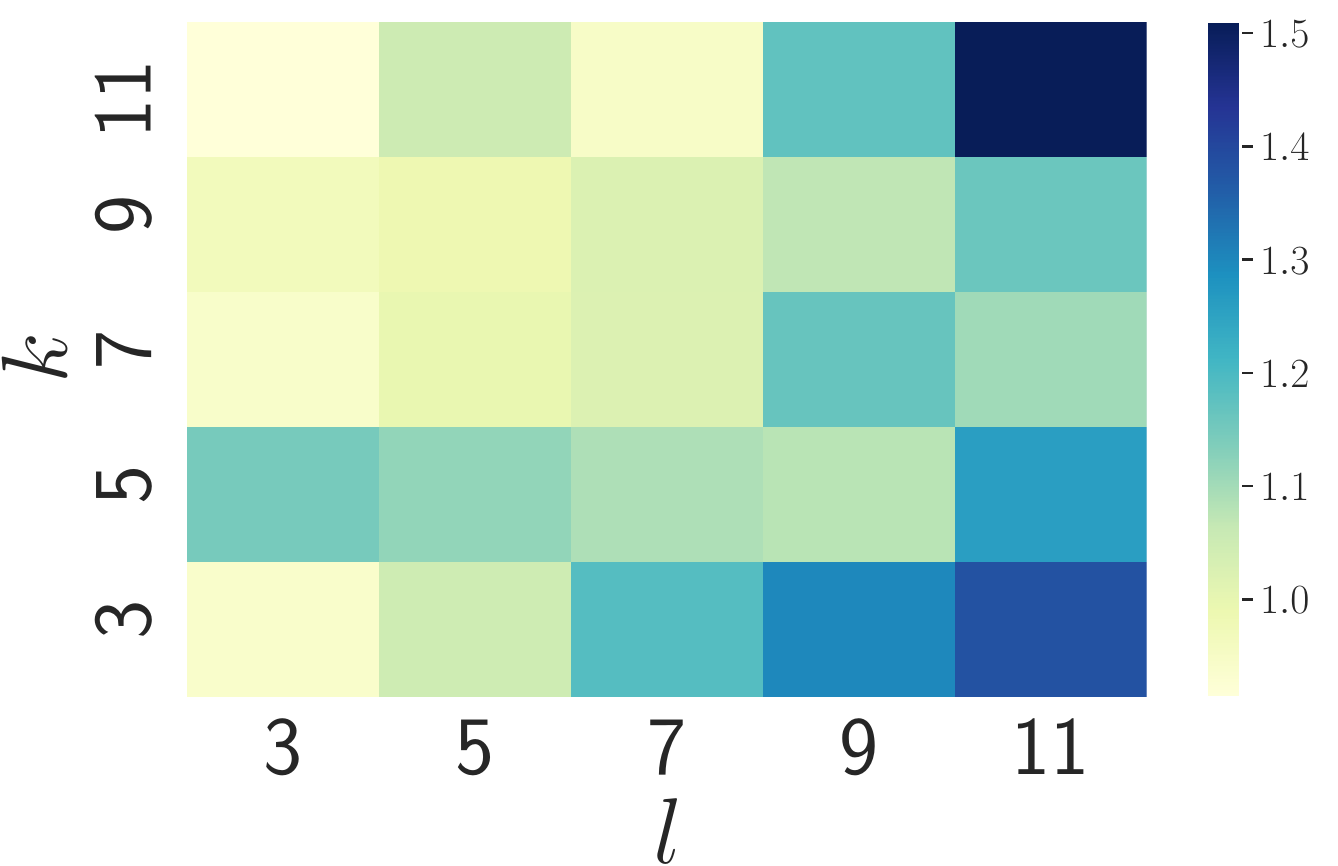}
		}%
		\centering
    \vskip -0.1in
		\caption{Hyper-parameter analysis: The results in terms of RMSE achieved by BroadCF with different ${k}$ and $l$.}
		\label{figure: num_neighbor_RMSE}
		\vskip -0.1in
	\end{figure*}
	
	\begin{figure*}[!t]
		\centering
\centerline{
		\subfigure[{ml-la}]{
			\centering
			\includegraphics[width=0.24\linewidth]{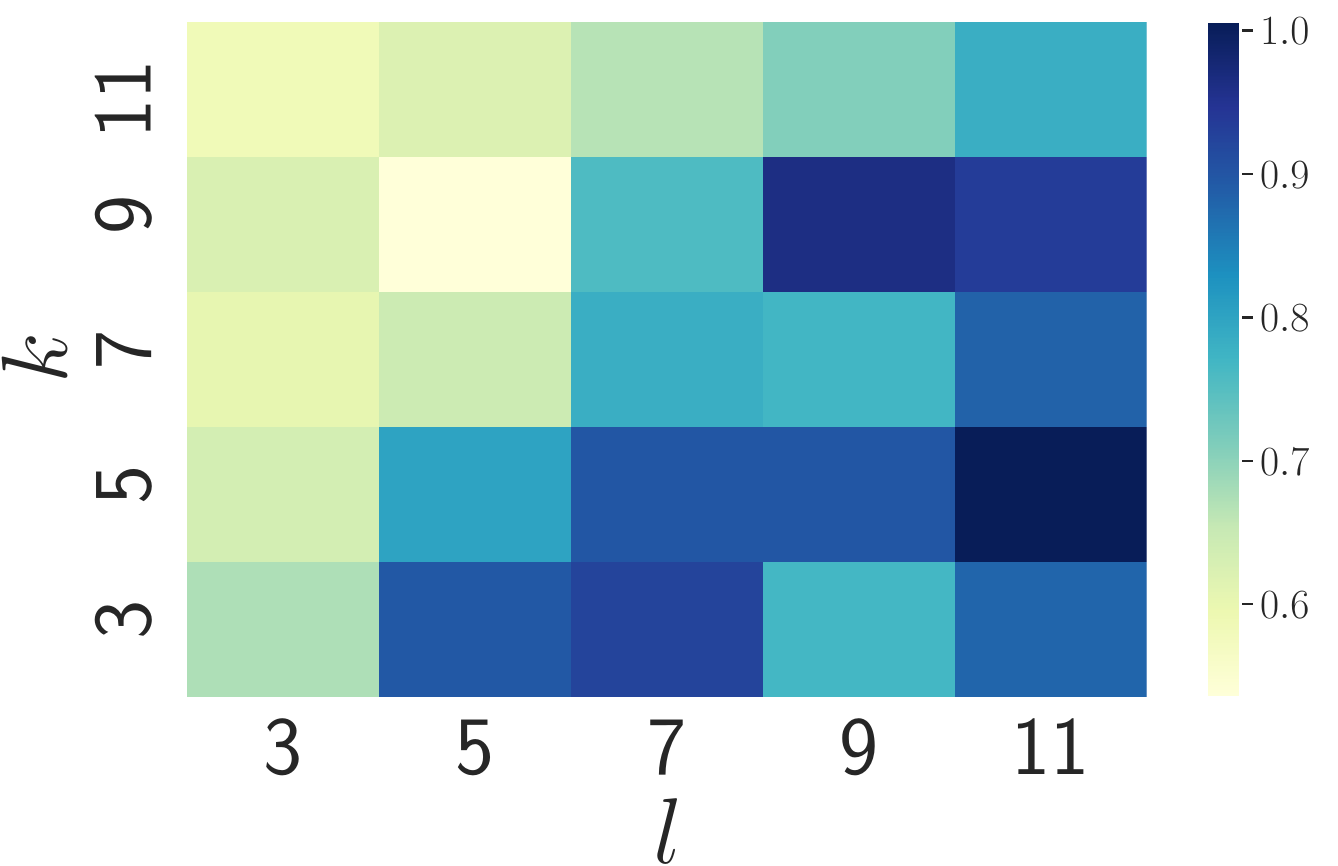}
		}%
		\subfigure[{Amazon-DM}]{
			\centering
			\includegraphics[width=0.24\linewidth]{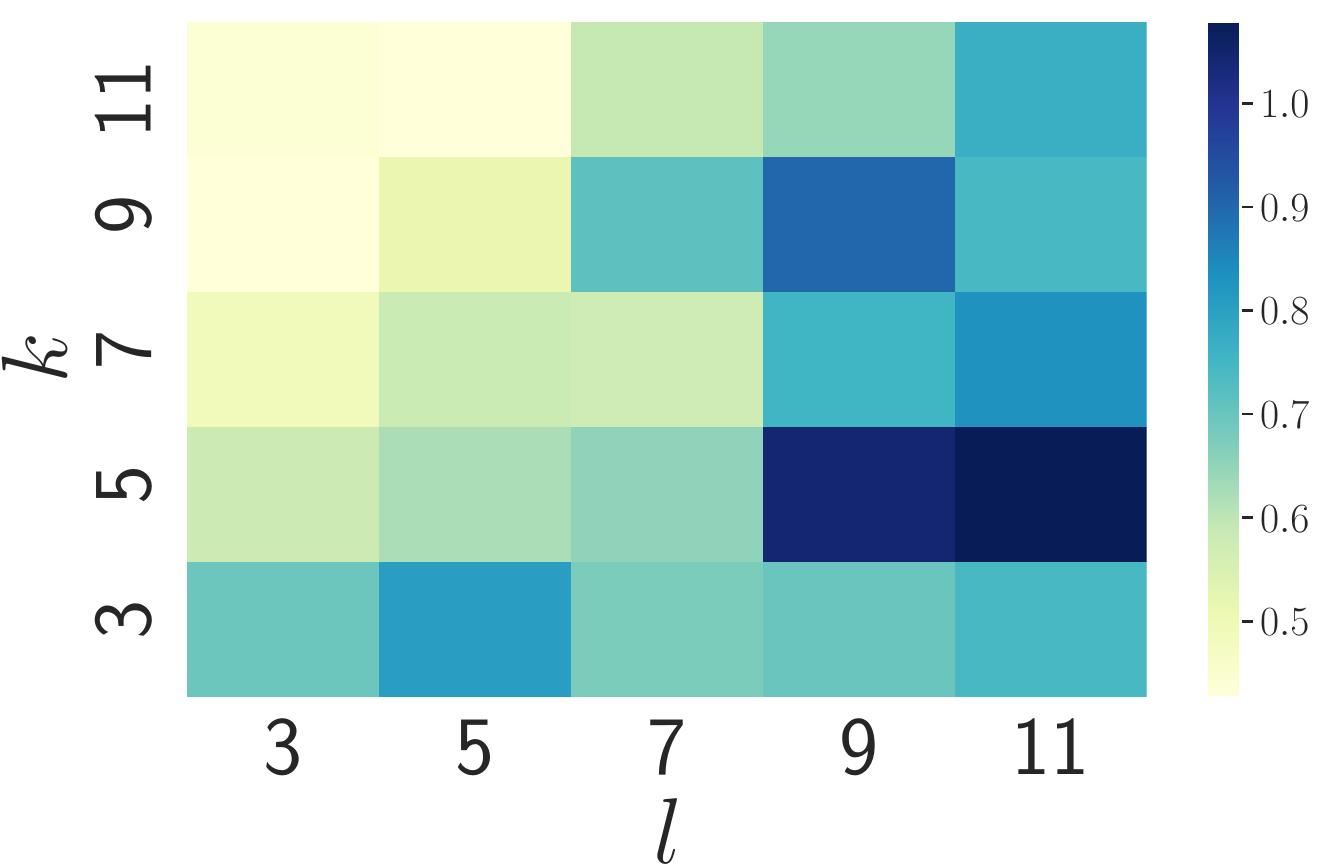}
		}%
		\subfigure[{Amazon-GGF}]{
			\centering
			\includegraphics[width=0.24\linewidth]{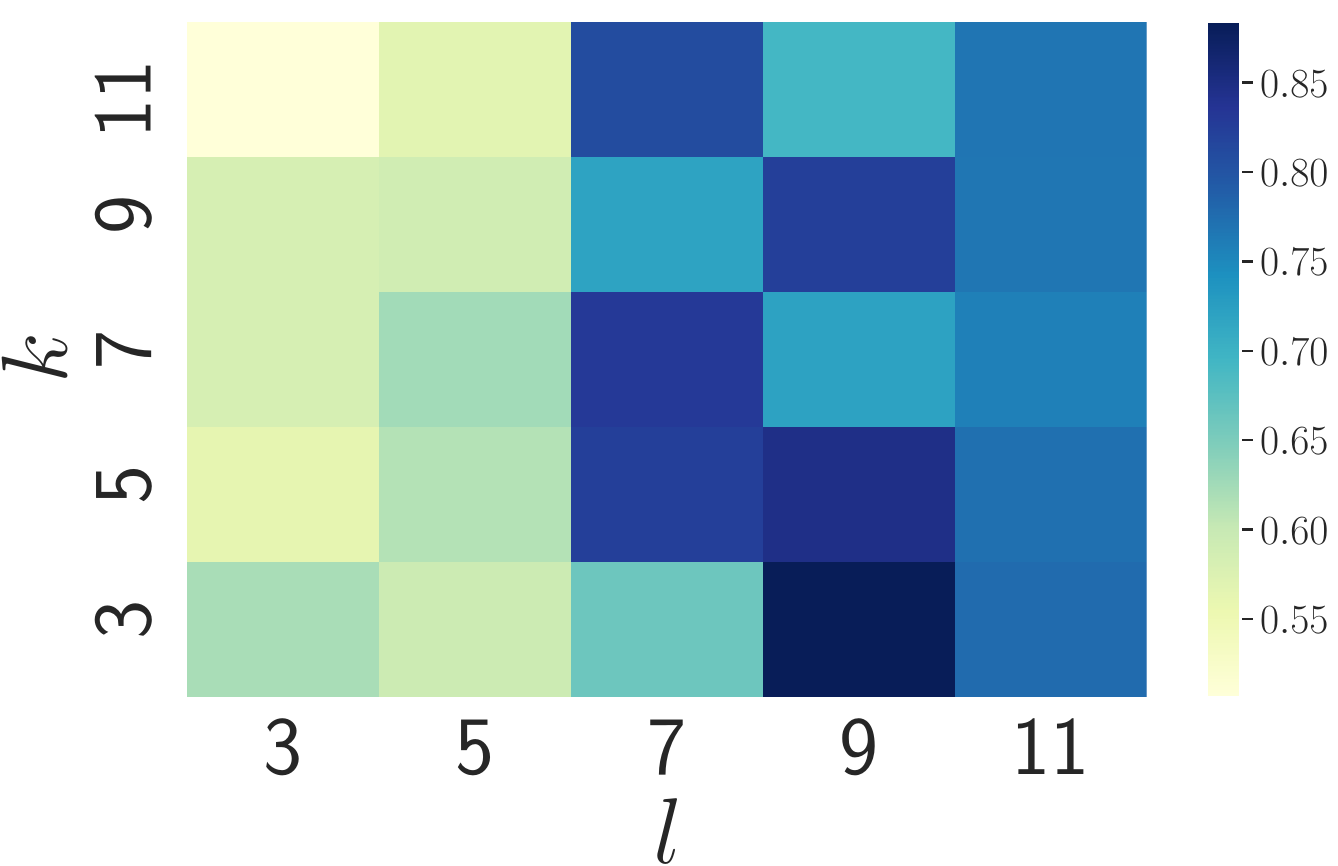}
		}
		\subfigure[{Amazon-PLG}]{
			\centering
			\includegraphics[width=0.24\linewidth]{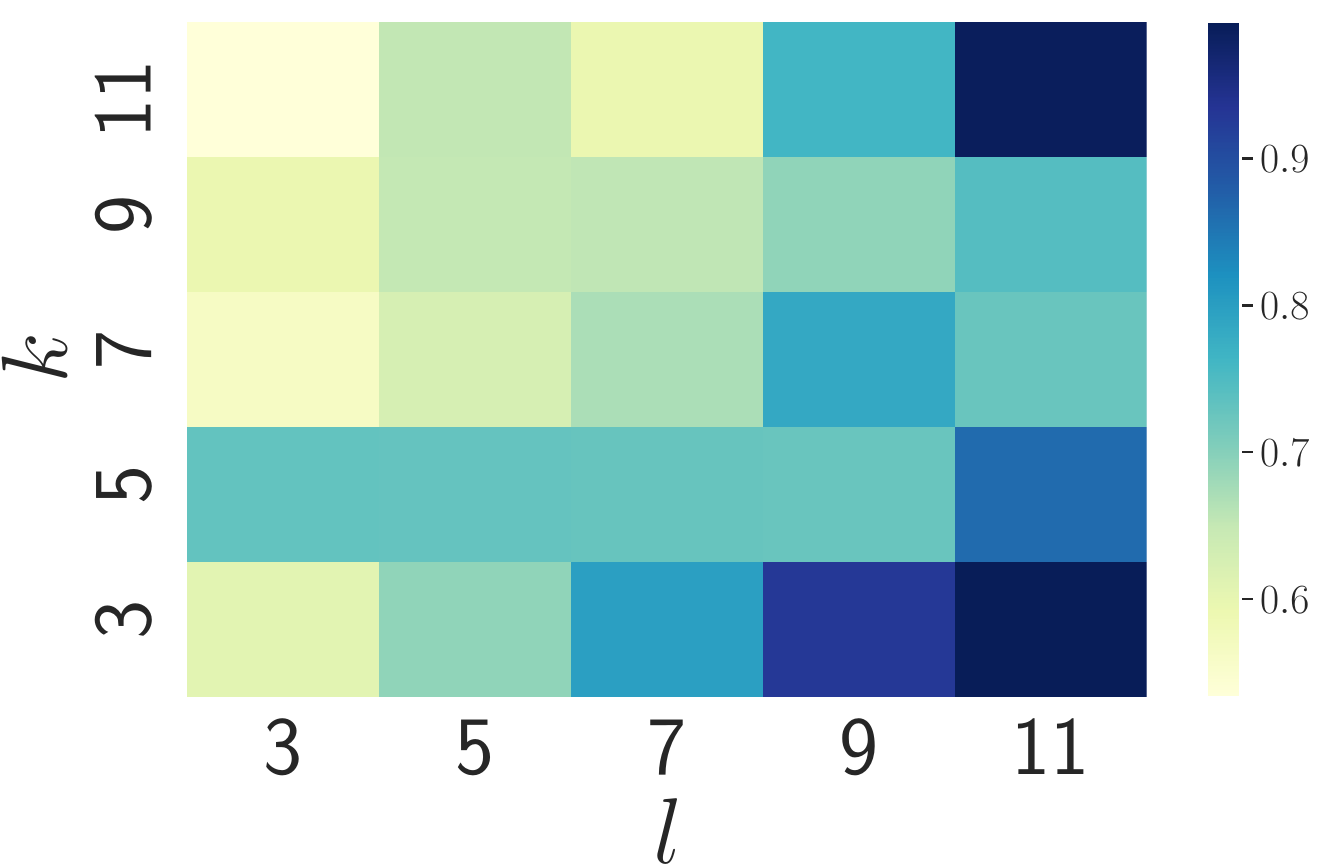}
		}%
}
		\centering
    \vskip -0.1in
		\caption{Hyper-parameter analysis: The results in terms of MAE achieved by BroadCF with different ${k}$ and $l$.}
		\label{figure: num_neighbor_MAE}
		\vskip -0.1in
	\end{figure*}
	
	\subsubsection{Comparison on Running Cost}

	Apart from the prediction performance, BroadCF has superiority on running cost. To this end, this subsection compares the running cost of BroadCF with the baselines, including the training time, testing time, and trainable parameter storage cost.

		\begin{figure*}[!t]
		\centering
\centerline{
		\subfigure[{ml-la}]{
			\centering
			\includegraphics[width=0.24\linewidth]{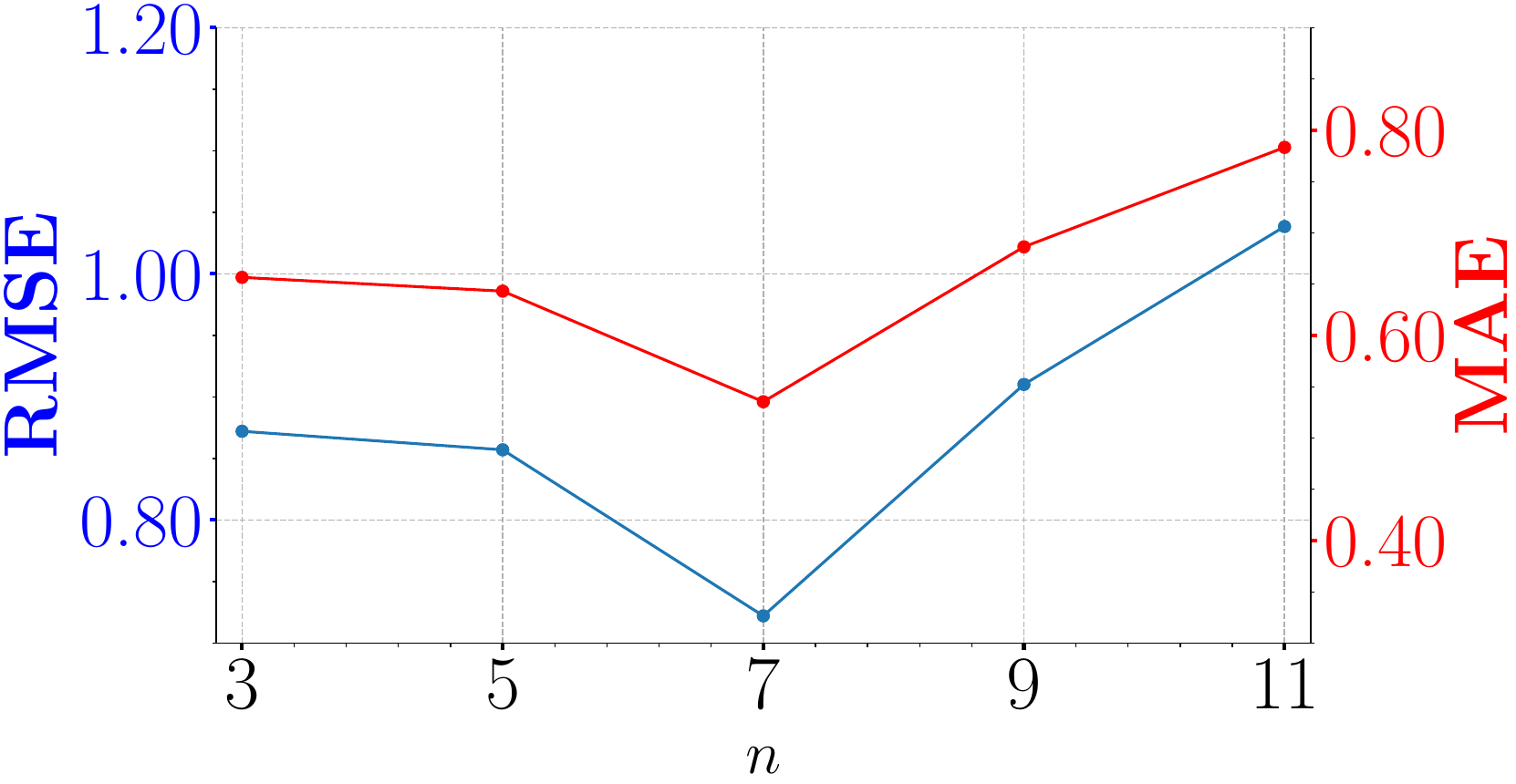}
		}%
		\subfigure[{Amazon-DM}]{
			\centering
			\includegraphics[width=0.24\linewidth]{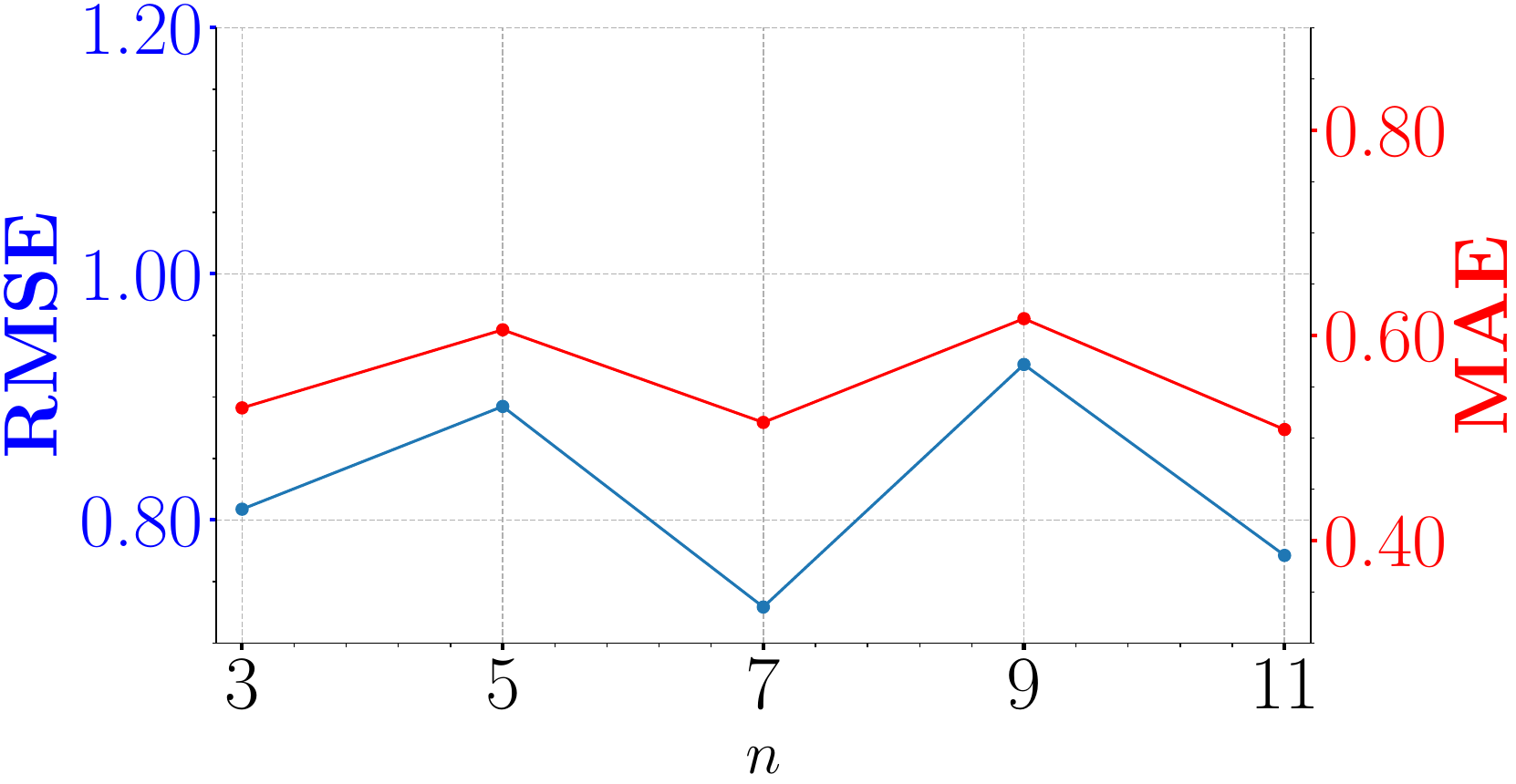}
		}%
		\subfigure[{Amazon-GGF}]{
			\centering
			\includegraphics[width=0.24\linewidth]{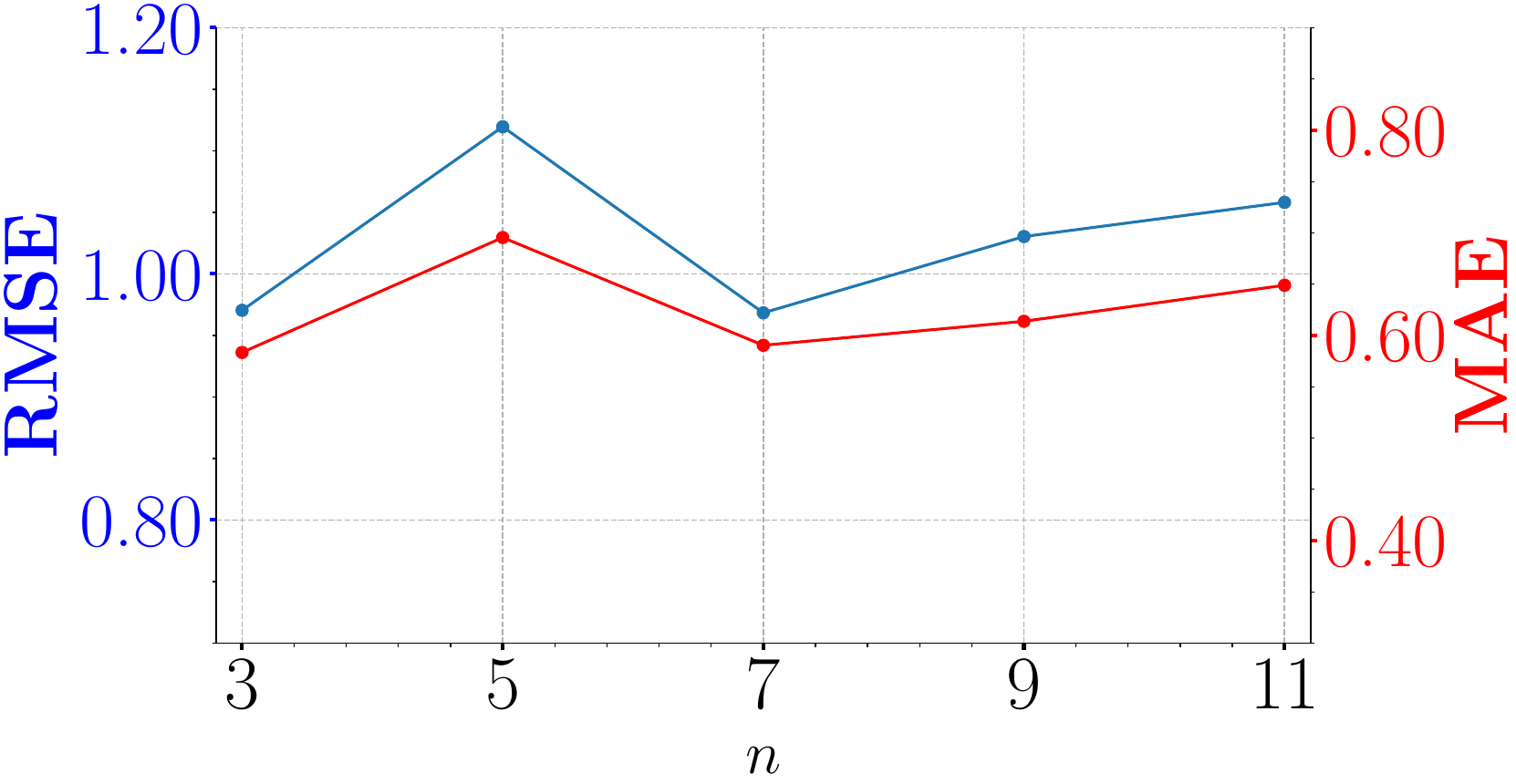}
		}
		\subfigure[{Amazon-PLG}]{
			\centering
			\includegraphics[width=0.24\linewidth]{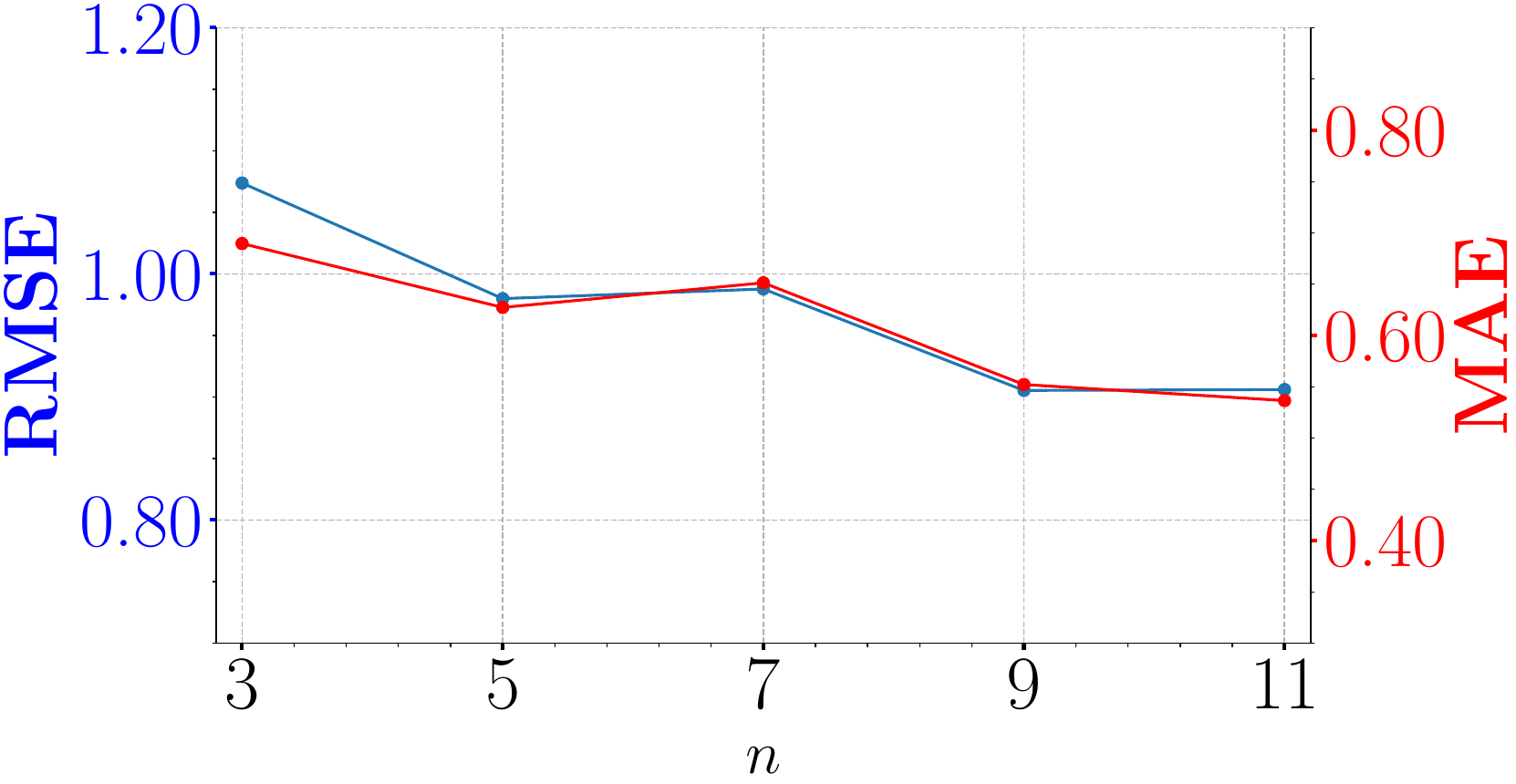}
		}%
}
		\centering
    \vskip -0.1in
		\caption{Hyper-parameter analysis: The results in terms of RMSE and MAE achieved by BroadCF with different $n$. }
		\label{Mapped Feature Groups}
  \vskip -0.1in
	\end{figure*}
	
	\begin{figure*}[!t]
		\centering
\centerline{
		\subfigure[{ml-la}]{
			\centering
			\includegraphics[width=0.24\linewidth]{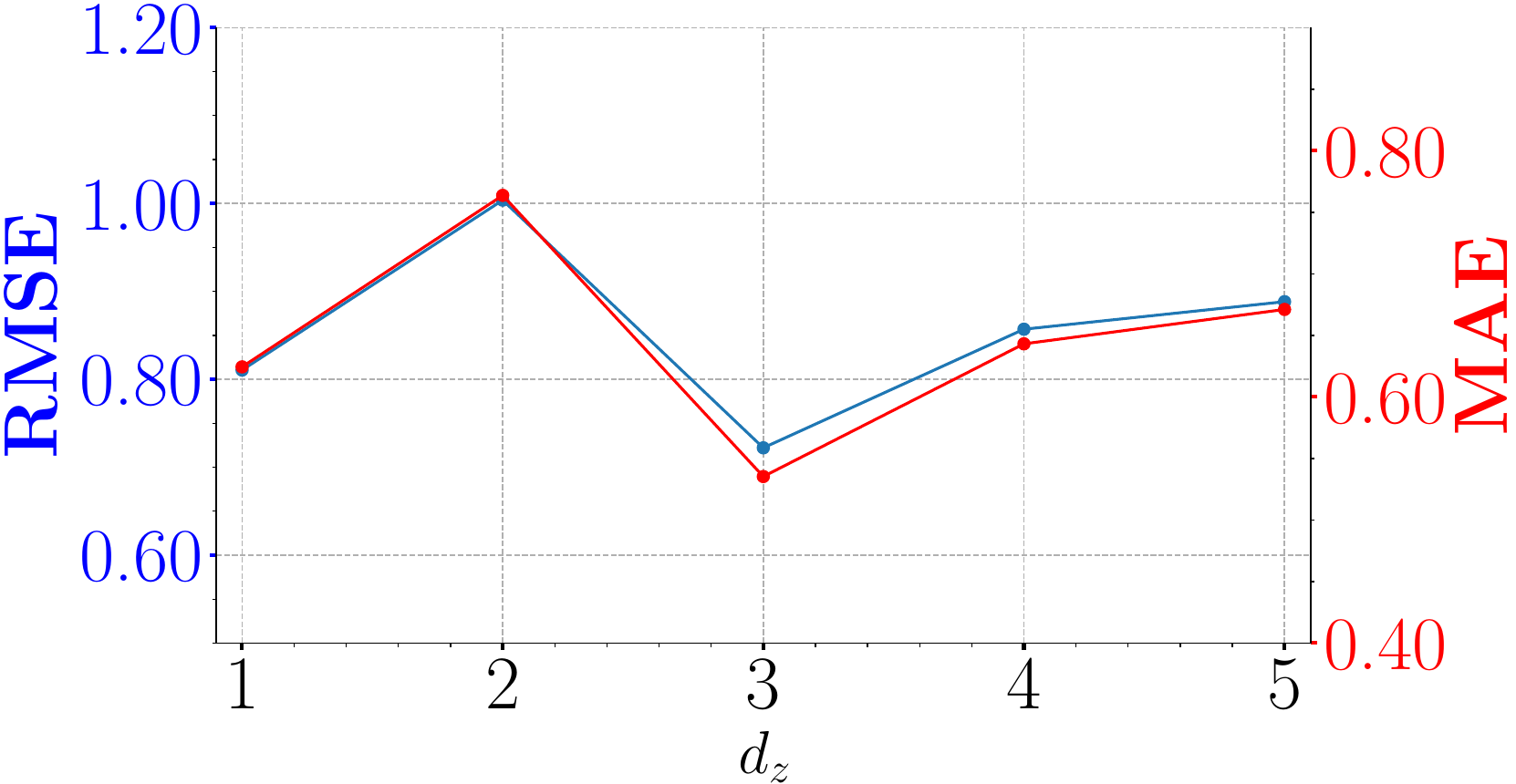}
		}%
		\subfigure[{Amazon-DM}]{
			\centering
			\includegraphics[width=0.24\linewidth]{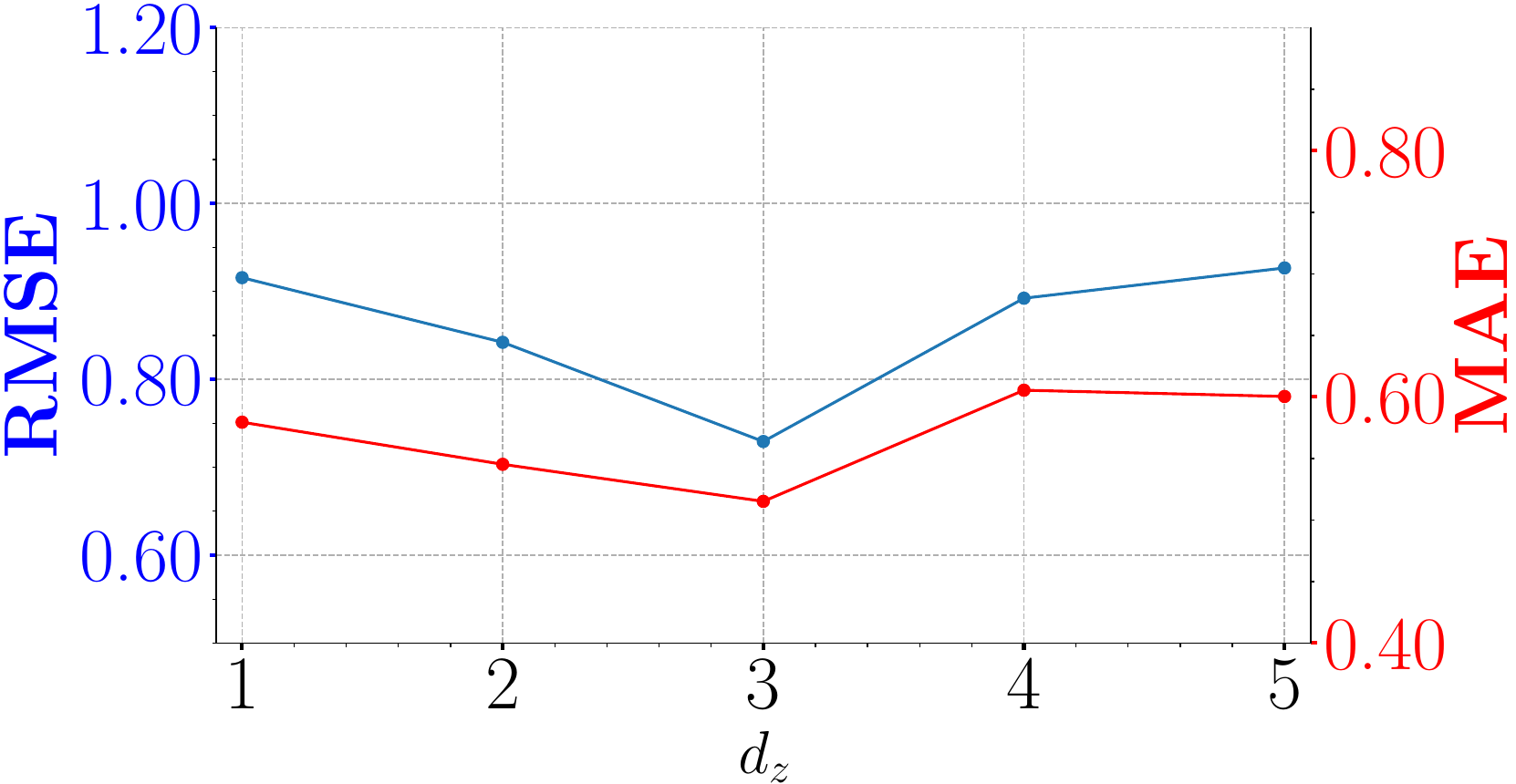}
		}%
		\subfigure[{Amazon-GGF}]{
			\centering
			\includegraphics[width=0.24\linewidth]{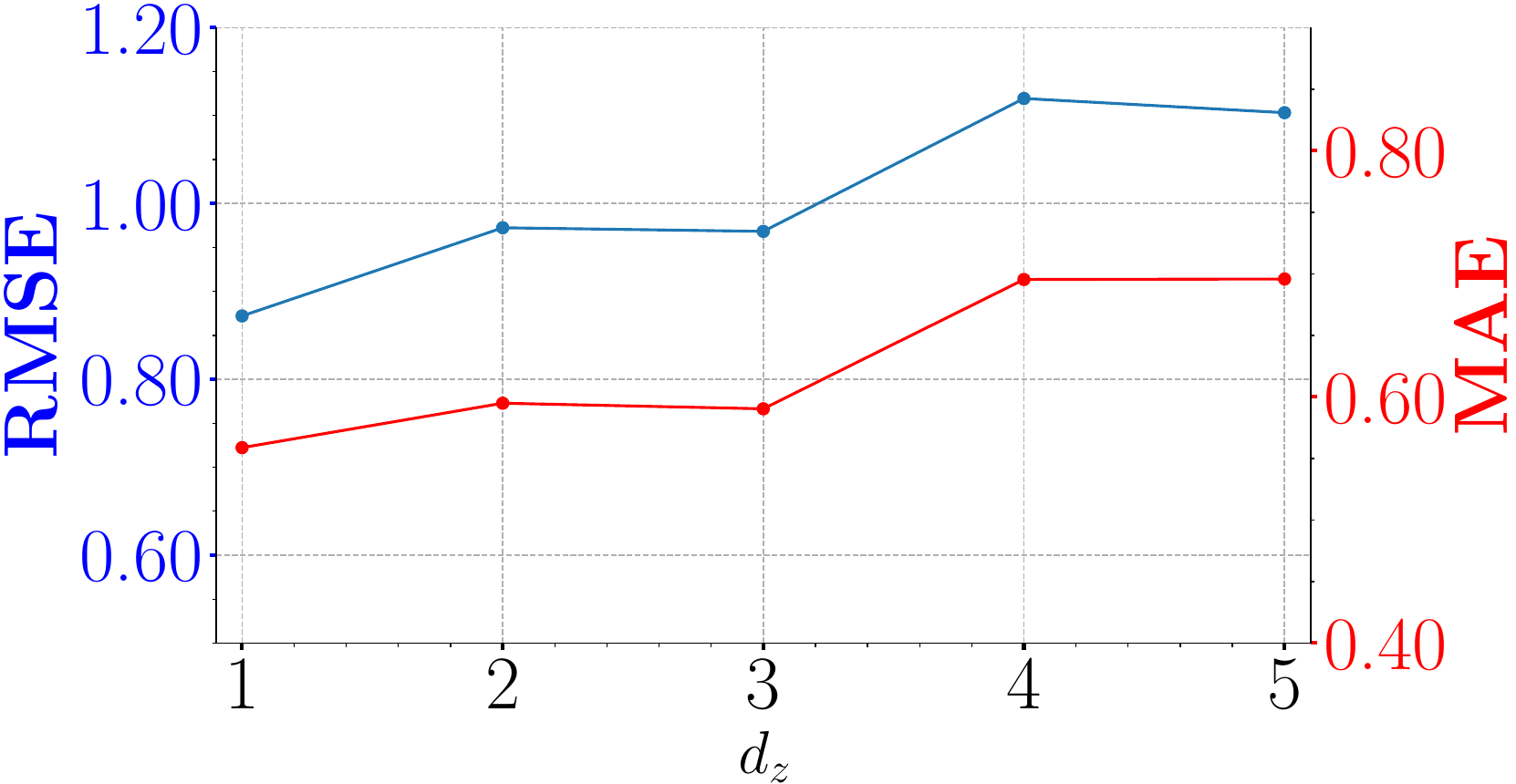}
		}
		\subfigure[{Amazon-PLG}]{
			\centering
			\includegraphics[width=0.24\linewidth]{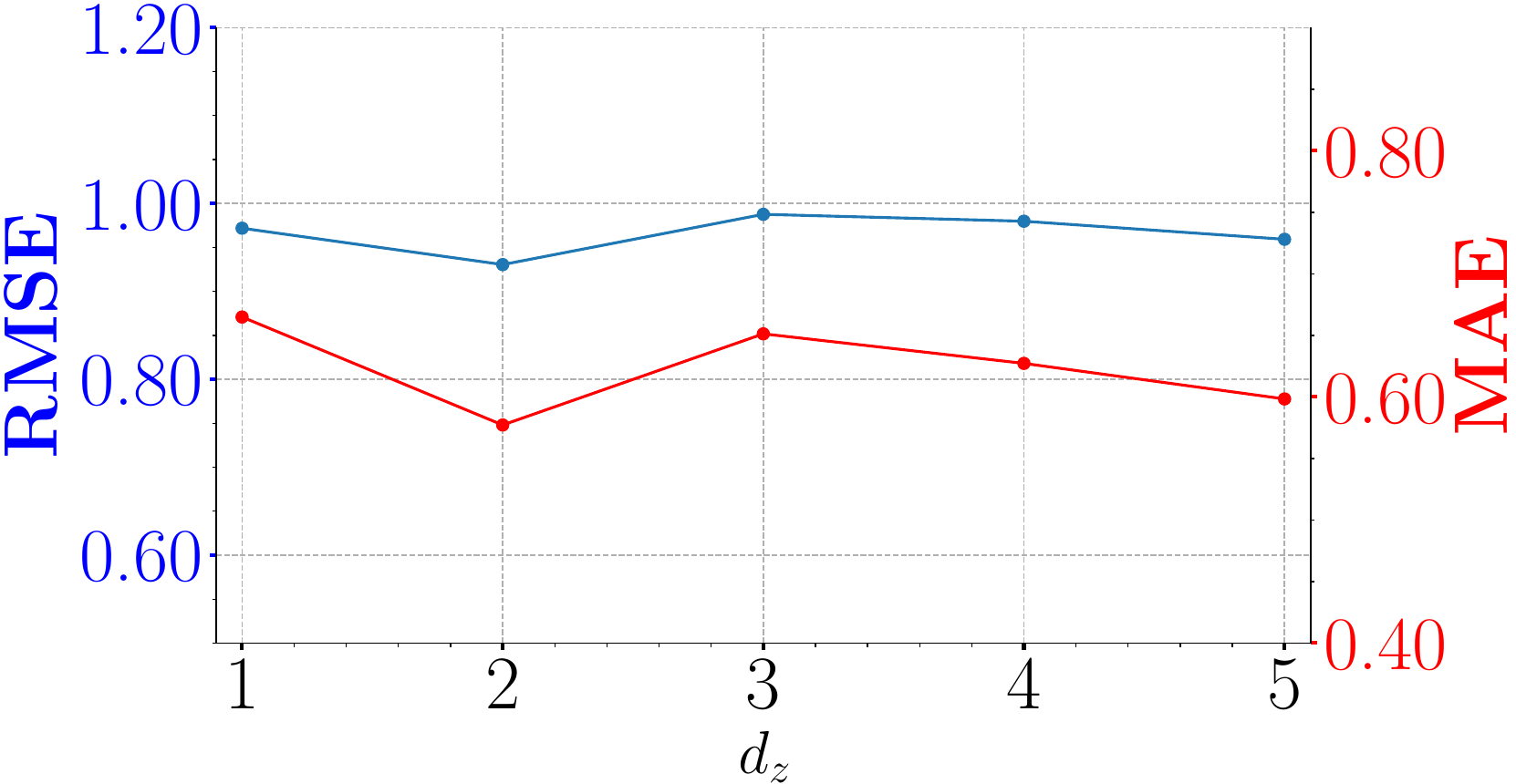}
		}%
}
		\centering
    \vskip -0.1in
		\caption{Hyper-parameter analysis: The results in terms of RMSE and MAE achieved by BroadCF with different $d_z$.}
		\label{Mapped Feature Dimension}
  \vskip -0.1in
	\end{figure*}

	Table~\ref{tab:training time} and Table~\ref{tab:testing time} report the training time and the testing time (in seconds) consumed by each algorithm on each dataset. Specifically, the training time and the testing time account for the time from feeding the original data into the model to outputting the results. According to the two tables, we can see that the proposed BroadCF algorithm is very efficient on running time. In particular, from Table~\ref{tab:training time}, BroadCF consumes the least training time among all methods on most datasets except Amazon-IV. On the Amazon-DM dataset, BroadCF is four times faster than the fastest baseline namely SHT on training time, and it consumes about 3.30\% training time by BPAM. On the Amazon-IV dataset, although BroadCF is slower than NeuMF, it is still faster than BPAM. Overall, from the perspective of training time, BroadCF has superiority compared with the baselines. This may be due to that the optimization problem of BroadCF is solved by using the ridge regression approximation of pseudoinverse without need of iterative training process.

From Table~\ref{tab:testing time}, we can see that although BroadCF is not the fastest algorithm on the testing time, it is the fastest among all the neural network based methods on most datasets except Amazon-IV. And the fastest method in terms of testing time is PMF. This is mainly due to that, in the testing procedure, BroadCF requires calculating the mapped features and enhanced features, in addition to the multiplication by the trainable weight vector, while PMF only requires to compute the matching score of the user/item latent factors. The above comparison results on training time and testing time have demonstrated the efficiency of the proposed BroadCF method from the perspective of running time.
	
		\begin{figure*}[!t]
		\centering
\centerline{
		\subfigure[{ml-la}]{
			\centering
			\includegraphics[width=0.24\linewidth]{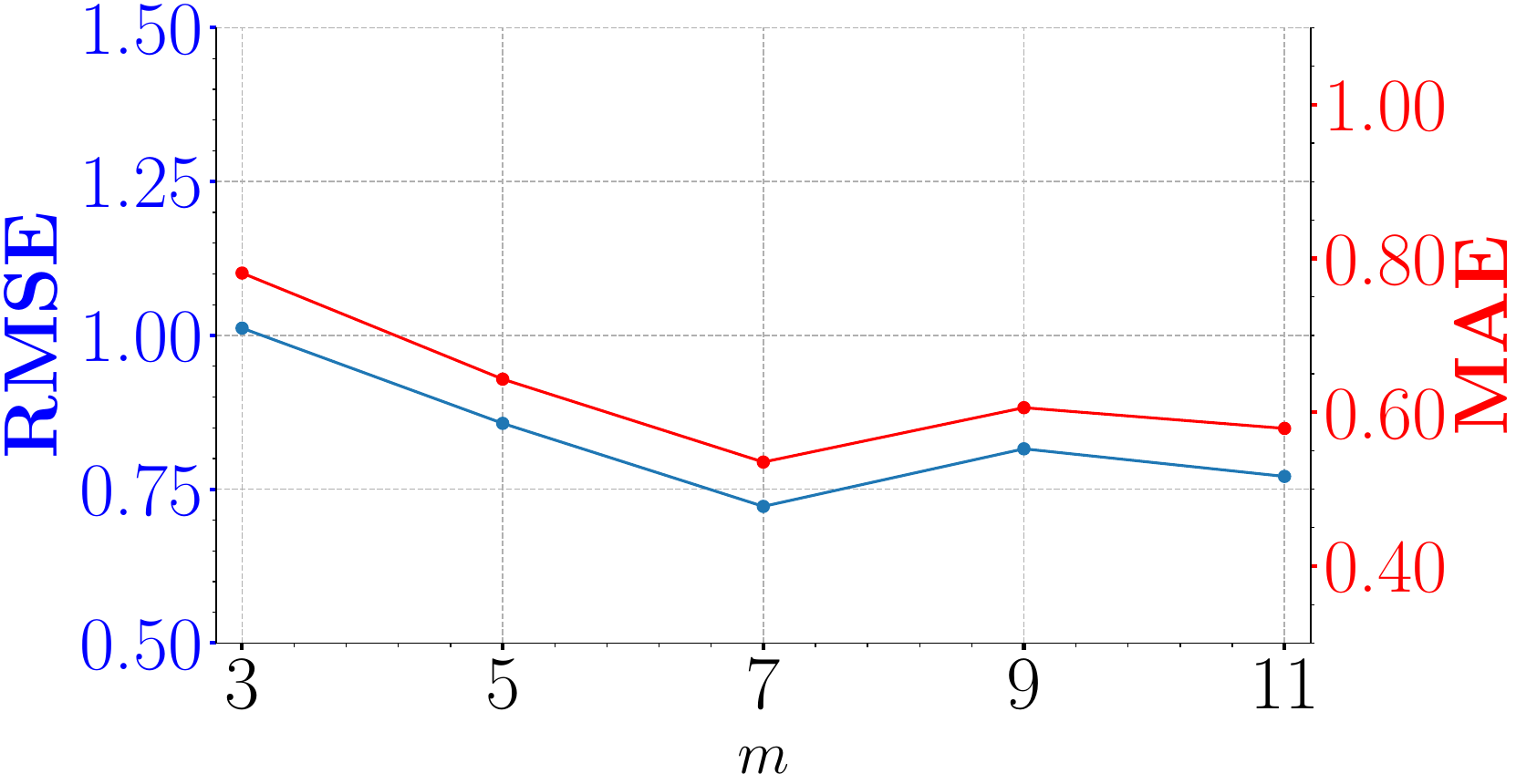}
		}%
		\subfigure[{Amazon-DM}]{
			\centering
			\includegraphics[width=0.24\linewidth]{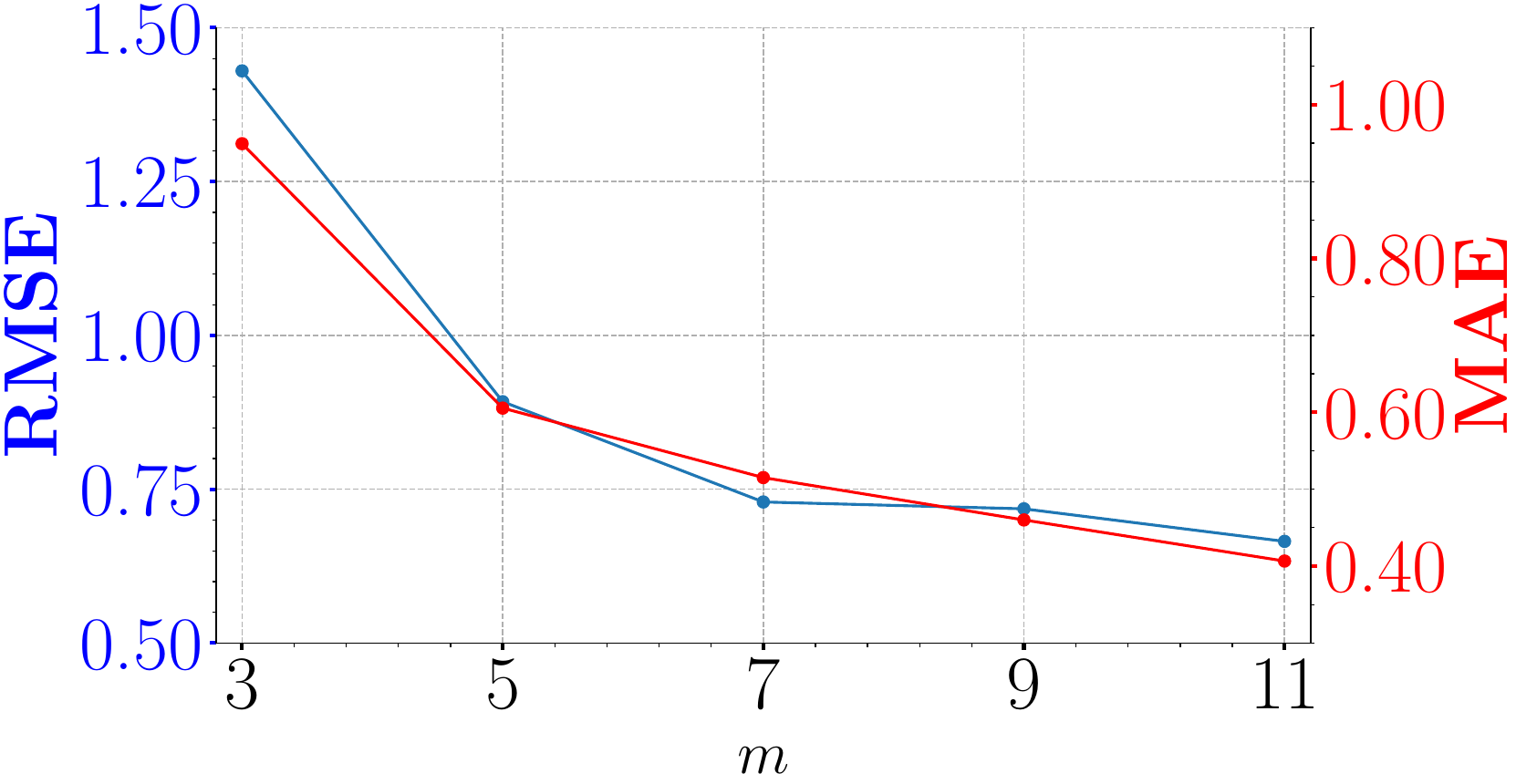}
		}%
		\subfigure[{Amazon-GGF}]{
			\centering
			\includegraphics[width=0.24\linewidth]{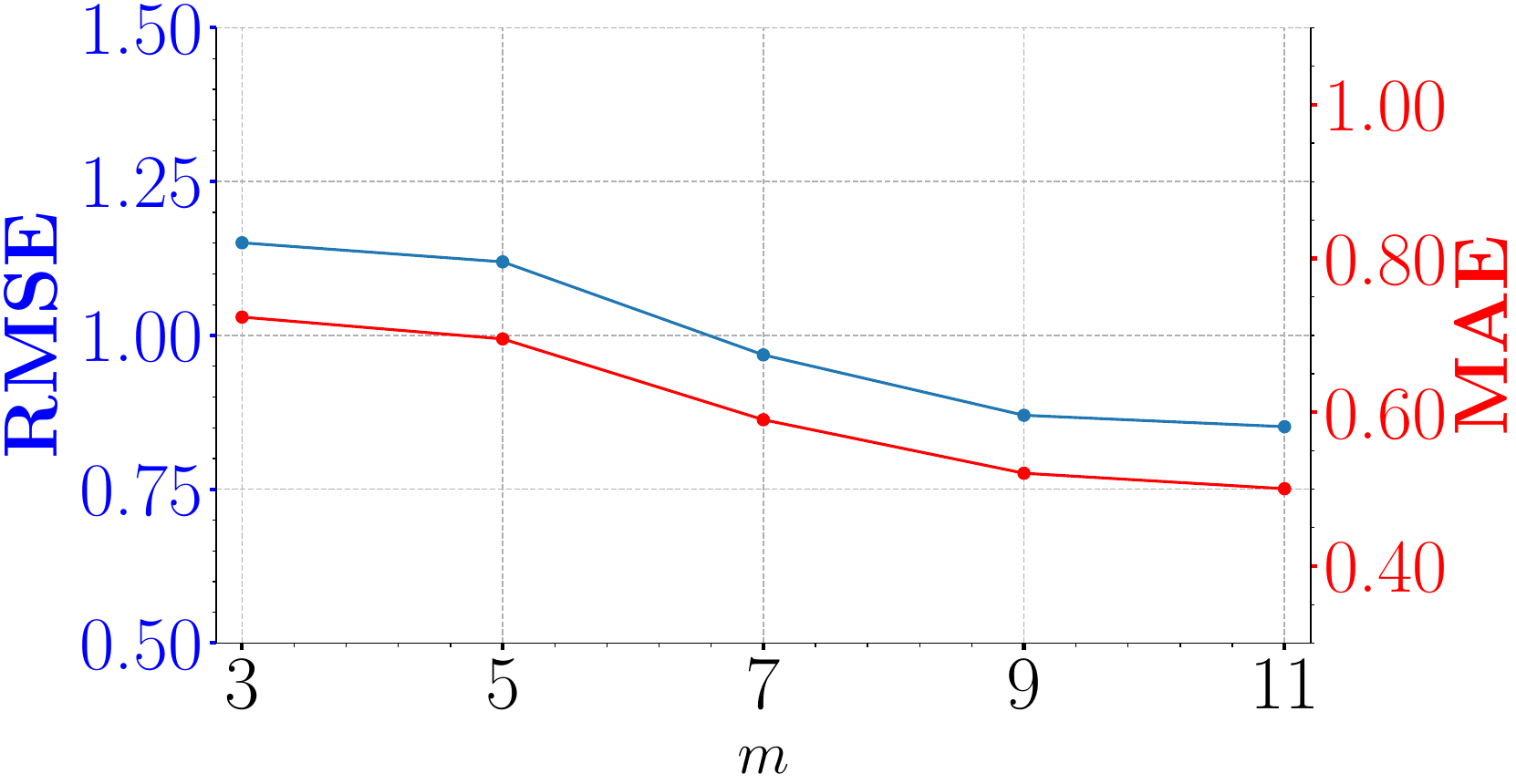}
		}%
		\subfigure[{Amazon-PLG}]{
			\centering
			\includegraphics[width=0.24\linewidth]{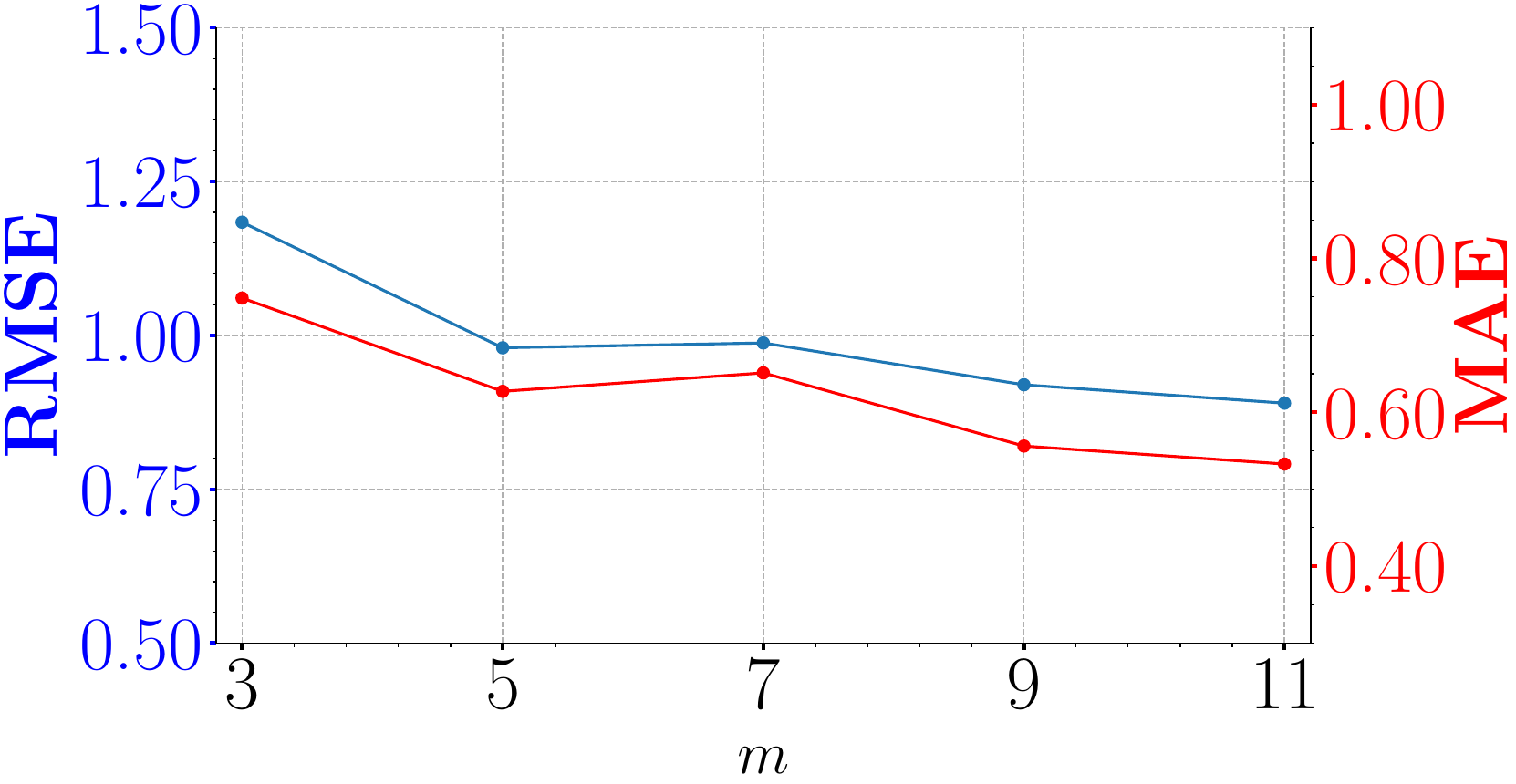}
		}%
}
  \vskip -0.1in
		\caption{Hyper-parameter analysis: The results in terms of RMSE and MAE achieved by BroadCF with different $m$.}
		\label{Enhanced Feature Groups}
  \vskip -0.1in
	\end{figure*}
	
	\begin{figure*}[!t]
		\centering
		\subfigure[{ml-la}]{
			\centering
			\includegraphics[width=0.24\linewidth]{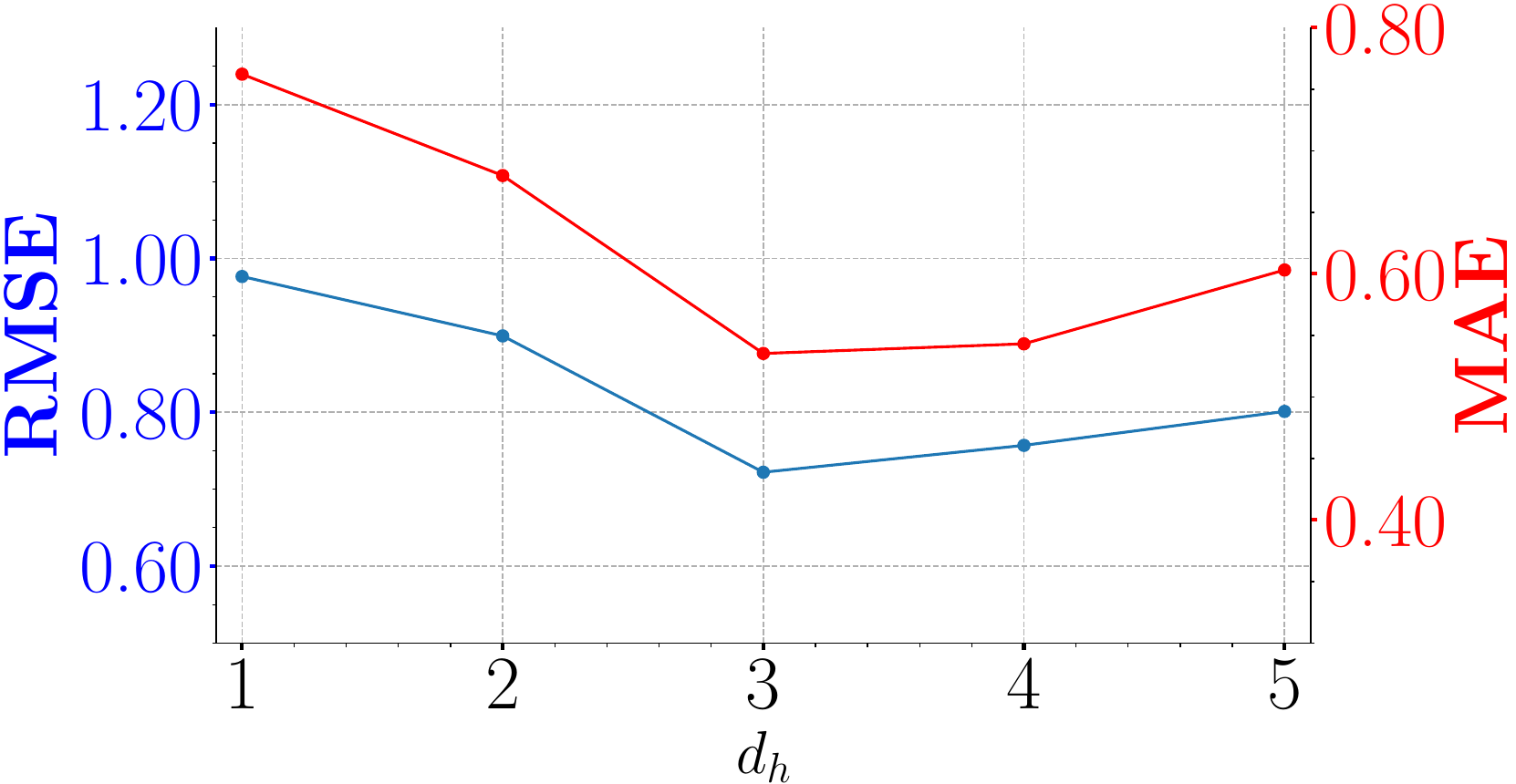}
		}%
		\subfigure[{Amazon-DM}]{
			\centering
			\includegraphics[width=0.24\linewidth]{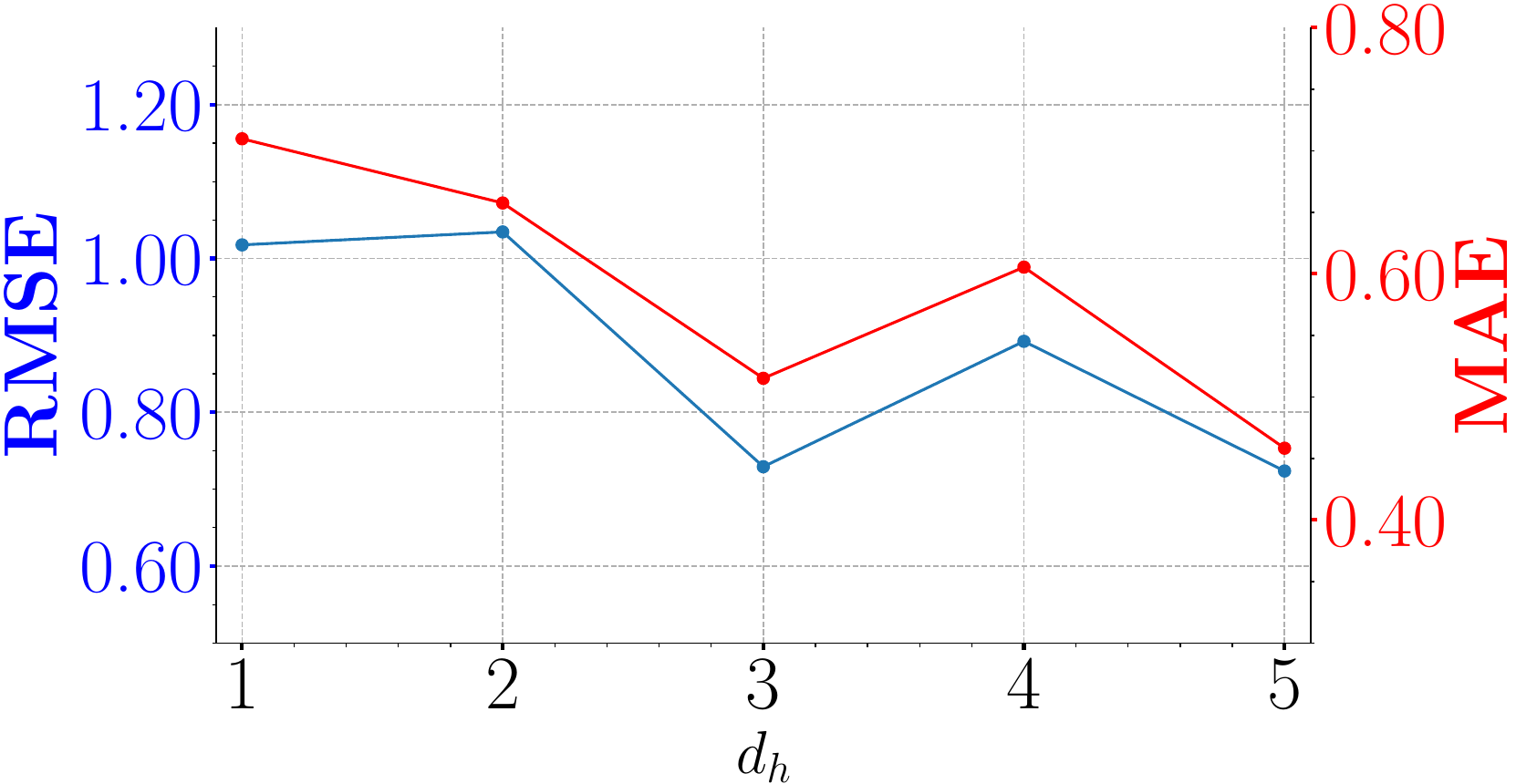}
		}%
		\subfigure[{Amazon-GGF}]{
			\centering
			\includegraphics[width=0.24\linewidth]{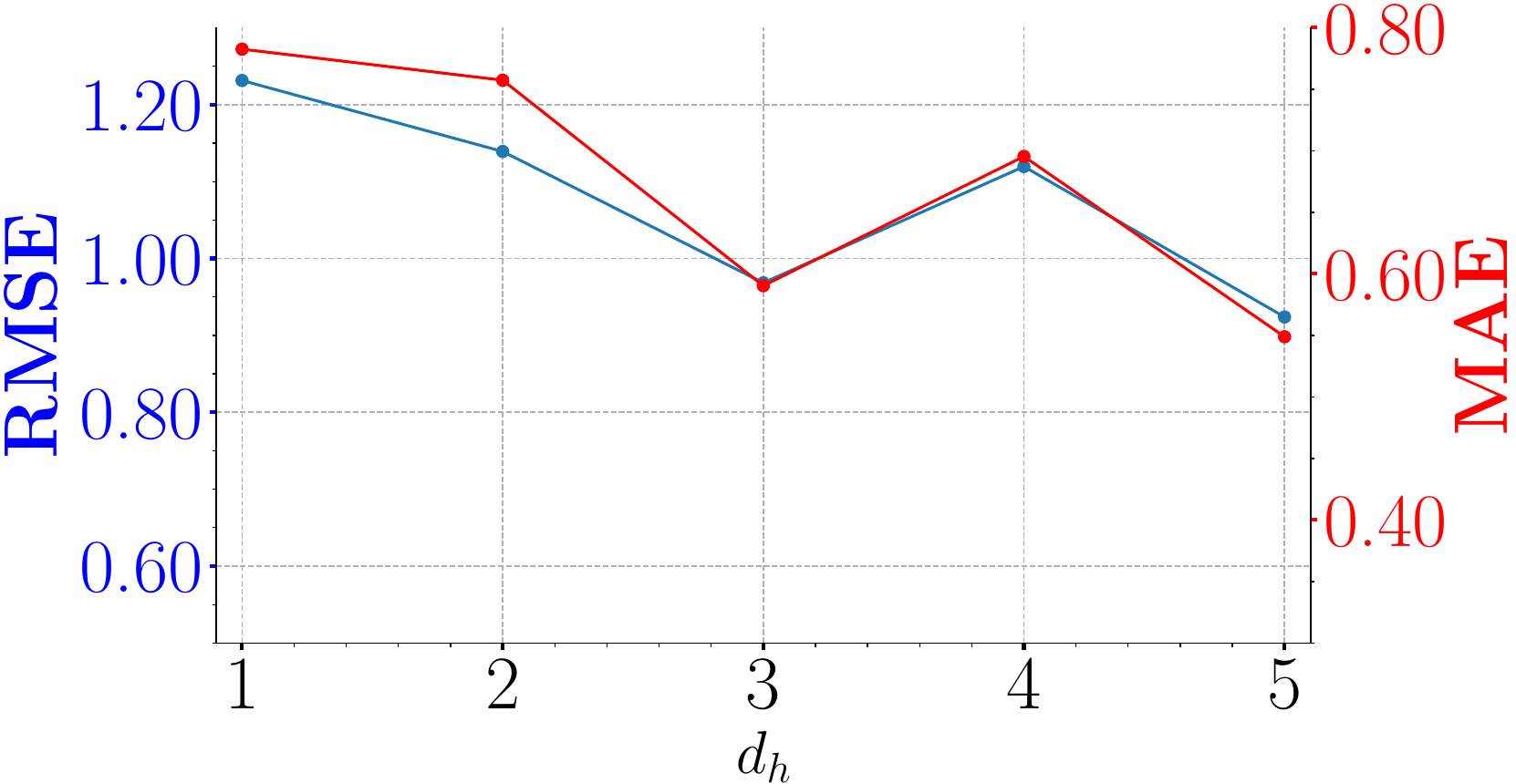}
		}%
		\subfigure[{Amazon-PLG}]{
			\centering
			\includegraphics[width=0.24\linewidth]{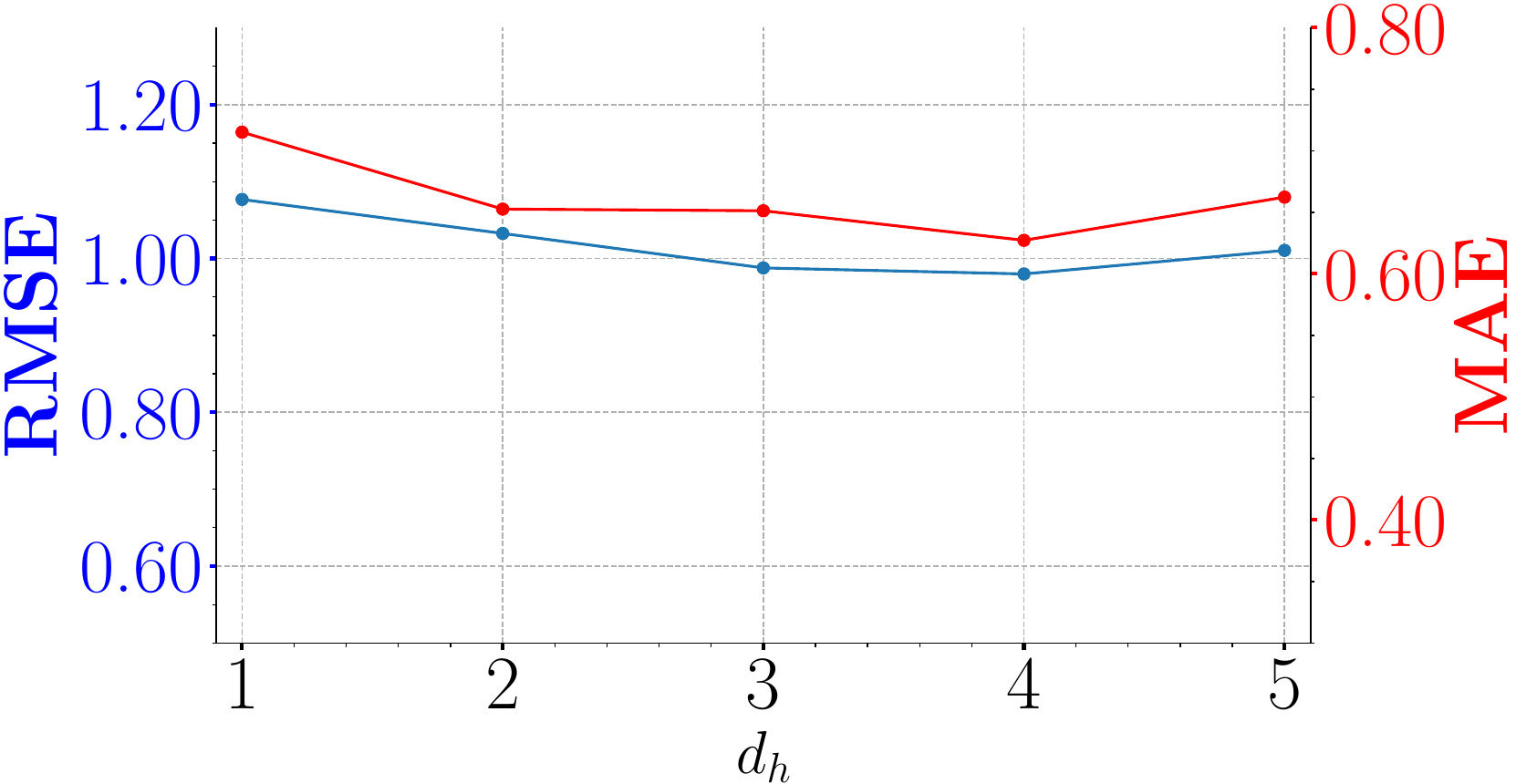}
		}%
		\centering
    \vskip -0.1in
		\caption{Hyper-parameter analysis: The results in terms of RMSE and MAE achieved by BroadCF with different $d_h$.}
		\label{Enhanced Feature Dimension}
  \vskip -0.1in
	\end{figure*}
	
	We also demonstrate the efficiency of BroadCF from the perspective of storage requirements. Following~\cite{DBLP:journals/tkde/WangXHZHL22}, Table~\ref{tab:storage costs} reports the number of trainable parameters to be stored during the training procedure of each algorithm on each dataset. From the table, we can see that, the proposed BroadCF algorithm is several-magnitude more efficient than all the baselines. This is mainly due to that the only trainable parameters to be stored in BroadCF are the weight vector $\mathbf{w}\in\mathbb{R}^{(nd_z+md_h)\times 1}$, which is independent of the input data size and dimension. On the contrary, all the baselines need to store much more trainable parameters, especially for the DNNs-based models. This result also confirms the fact that BroadCF requires several-magnitude less training parameters, even less than the linear model namely PMF. Despite the relatively small number of trainable parameters, BroadCF can still well capture the nonlinear relationships in the user-item pairs and therefore generates much better results as shown in the comparison of prediction. This is mainly due to the advantage achieved by mapping the original features into the mapped features and then into the enhanced features via random matrices plus nonlinear activation functions, which can be regarded as learning nonlinear representations from different views. Overall, the above comparison results on running cost have confirmed the efficiency of the proposed BroadCF algorithm from the perspective of trainable parameter storage requirement.

	\subsection{Hyper-Parameter Analysis}
	In this subsection, we will analyze the effect of six hyper-parameters on the prediction performance of BroadCF, \textit{i.e.}, nearest user number $k$, nearest item number $l$, number of mapped feature groups $n$, mapped feature dimension $d_z$, number of enhanced feature groups $m$, and enhanced feature dimension $d_h$. When investigating one hyper-parameter, the other hyper-parameters are fixed. Due to the space limit, only the first four datasets are analyzed in this part of experiments, i.e., ml-la, Amazon-DM, Amazon-GGF and Amazon-PLG. Notice that the three Amazon datasets cover different degrees of sparsity, namely 99.90\%, 99.96\% and 99.98\%. And similar observations can be obtained on the remaining datasets.
	
	\subsubsection{Number of the Nearest Users/Items}
	
		First of all, we will investigate the effect of the number of the nearest users/items, namely $k$ and $l$, on the prediction performance of BroadCF. To this end, we run BroadCF by setting $k$ and $l$ in the range of [3, 5, 7, 9, 11], and report the prediction performance measured by RMSE and MAE in \figurename~\ref{figure: num_neighbor_RMSE} and \figurename~\ref{figure: num_neighbor_MAE} respectively. From the figures, we can observe that a relatively stable performance is achieved when using different $k$ and $l$ on most datasets. That is, the variances of the RMSE and MAE values in the two figures are relatively small as shown in the heatmaps. However, a general trend is that when increasing $k$ and $l$, the prediction performance would slightly degenerate expect on the ml-la dataset. This is mainly due to the long-tail distribution, \textit{i.e.}, most of users/items have very few ratings. Setting a relatively large number of the nearest users/items would mistakenly introduce more noises, \textit{i.e.}, filling the missing entries by means of the nearest rating via Eq.~(\ref{eq:barp}) and Eq.~(\ref{eq:barq}). And the exception on the ml-la dataset has confirmed this analysis. That is, ml-la is a relatively dense dataset compared with the other datasets, in which the users/items contain more ratings. Regarding the trade-off between the running efficiency and the accuracy, we set $k$ and $l$ to 5 in the experiments.

  \begin{table*}[!t]
	\caption{{Ablation {s}tudy on the {seven} datasets: Comparison between BroadCF and {its} {t}wo {v}ariants {n}amely BroadRS and MCF.} }
	\label{tab: ablation study}
 \vskip -0.1in
 	\centering
\begin{tabular}{lllllllll}
			\hline
			\multicolumn{1}{c}{\textbf{Methods}} &  \multicolumn{1}{c}{\textbf{{Metric}}} & \multicolumn{1}{c}{\textbf{ml-la}} & \multicolumn{1}{c}{\textbf{Amazon-DM}} & \multicolumn{1}{c}{\textbf{Amazon-GGF}}  & \multicolumn{1}{c}{\textbf{Amazon-PLG}} & \multicolumn{1}{c}{\textbf{Amazon-Automotive}} & \multicolumn{1}{c}{\textbf{{Amazon-Baby}}}
   & \multicolumn{1}{c}{\textbf{{Amazon-IV}}} \\\hline
\multirow{4}{*}{\textbf{BroadRS}} & RMSE & 1.1153 & 1.8020 & 2.6843 & 2.7709 & 2.8703 & 2.7525 & 0.9281 \\
                  & MAE & 0.8660 & 1.2475 & 1.2648 & 1.2236 & 1.0343  & 1.1185 & 0.6189\\
                  & NDCG@10 & 0.2048 & 0.1570 & 0.0915 & 0.1025 & 0.1182 & 0.0901 & 0.1757\\
                  & HR@10 & 0.3973 & 0.3076 & 0.3014 & 0.2777 & 0.2786  & 0.1797 & 0.4291\\ \hline
\multirow{4}{*}{\textbf{MCF}} & RMSE & 0.7689 & 0.7619 & 1.0435 & 1.0755 & 0.8755 & 0.9659 & 0.9238 \\
                  & MAE & 0.5520 & 0.5839 & 0.6165 & 0.7027 & 0.5395 & 0.6731 & 0.5588 \\
                  & NDCG@10 & 0.3647  & 0.1065 & 0.1703 & 0.0820 & 0.0747 & 0.1664 & 0.1204 \\
                  & HR@10 & 0.5950 & 0.2142 & 0.4505 & 0.3572 & 0.2926 & 0.3530 & 0.4208\\ \hline
\multirow{4}{*}{\textbf{BroadCF}} & RMSE & 0.7220 & 0.7292 & 0.9683 & 0.9877 & 0.8570 & 0.8528  & 0.7933\\
                  & MAE & 0.5353 & 0.5151 & 0.5903 & 0.6512 & 0.5202 & 0.5672 & 0.3994 \\
                  & NDCG@10 & 0.4765 & 0.3716 & 0.3484 & 0.3142 & 0.2767 & 0.2951 & 0.2699\\
                  & HR@10 & 0.8229 & 0.7927 & 0.7599 & 0.6541 & 0.7927 &  0.5792 & 0.6439 \\ \hline
\end{tabular}
\vskip -0.1in
\end{table*}
	
	\subsubsection{Hyper-Parameters in Feature Mapping}

		Secondly, we will investigate the effect of the hyper-parameters in the feature mapping module, including the number of mapped feature groups $n$ and the mapped feature dimension $d_z$. In the proposed BroadCF algorithm, the feature mapping module plays the role of mapping the original input matrix composed of user-item collaborative vectors into the mapped feature matrix, which is the first step of capturing the nonlinear relationships in the user-item pairs. The two hyper-parameters would affect the performance of this mapping capability.
	To this end, we run BroadCF by setting $n$ and $d_z$ in the ranges of [3,5,7,9,11] and [1,2,3,4,5] respectively, and report the RMSE and MAE values as a function of $n$ and $d_z$  in \figurename~\ref{Mapped Feature Groups} and \figurename~\ref{Mapped Feature Dimension} respectively. From the two figures, although the best values of $n$ and $d_z$ differ on different datasets, the performance variation is relatively small. This confirms the relative insensitivity of the proposed BroadCF algorithm to the hyper-parameters in feature mapping.  And in our experiments, we set $n$ and $d_z$ to 7 and 3 respectively.

	\subsubsection{Hyper-Parameters in Feature Enhancement}

		Similar to the previous subsection, we will also investigate the effect of the hyper-parameters in feature enhancement module, including the number of enhanced feature groups $m$ and the enhanced feature dimension $d_h$. In the proposed BroadCF algorithm, the feature enhancement module plays the role of mapping the mapped feature matrices into the enhanced feature matrix, which is the second step of capturing the nonlinear relationships in the user-item pairs. The two hyper-parameters would affect the prediction performance of this enhancement capability.
	Similarly, we run BroadCF by setting $m$ and $d_h$ in the ranges of [3,5,7,9,11] and [1,2,3,4,5] respectively, and report the RMSE and MAE values as a function of $m$ and $d_h$  in \figurename~\ref{Enhanced Feature Groups} and \figurename~\ref{Enhanced Feature Dimension} respectively. From the two figures, although the best values of $m$ and $d_h$ differ on different datasets, the performance variation is relatively small. This confirms the relative insensitivity of the proposed BroadCF algorithm to the hyper-parameters in feature enhancement. And in our experiments, we set $m$ and $d_h$ to 7 and 3 respectively.

\subsection{{Ablation Study}}
{Table~\ref{tab: ablation study} reports the results of the ablation study on the {seven} datasets. Specifically, we set up two variants of BroadCF for each dataset in order to investigate the effectiveness of the data preprocessing module and the BLS module. {The first variant is BroadRS without the data preprocessing module and the second variant is MCF using the MLP module instead of the BLS module.} The BroadCF model performs better than {the two variants} on all datasets, as shown in Table~\ref{tab: ablation study}. More specifically, the BroadRS variant operating without the data preprocessing module performs noticeably worse than both BroadCF and MCF, indicating that the data preprocessing module is essential. On the {ml-la} dataset, which has a sparsity of 98.30\%, we can observe that the improvement of MCF over BroadRS {in terms of} RMSE is 31.06\%, and the improvement of BroadCF over BroadRS {in terms of} RMSE is 35.26\%. Additionally, the improvement of BroadCF over BroadRS {in terms of} RMSE is 64.14\%, and the improvement of MCF over BroadRS {in terms of} RMSE is 61.18\% on the Amazon-PLG dataset, which has a 99.98\% sparsity. This indicates that even in cases where the dataset is excessively sparse, the proposed KNU Rating Vector and LNI Rating Vector still yield good results. The main reason for the superior performance of BroadCF over MCF on all {seven} datasets is {that the former can avoid the overfitting issue when taking  the low-dimensional KNU Rating Vector and LNI Rating Vector as input.}}

\section{{Conclusions and Future Work}}
\label{sec:Conclusion}
	
	In this paper, we have developed a novel neural network based recommendation method called Broad Collaborative Filtering (BroadCF). Compared with the existing DNNs-based recommendation methods, the proposed BroadCF method is also able to capture the nonlinear relationships in the user-item pairs, and hence can generate very satisfactory prediction performance. However, the superiority of BroadCF is that it is much more efficient than the DNNs-based methods, \textit{i.e.}, it consumes relatively very short training time and only requires relatively small amount of data storage, in particular trainable parameter storage. The main advantage lies in designing a data preprocessing procedure to convert the original rating data into the form of (Rating collaborative vector, Rating), which is then feed into the very efficient BLS for rating prediction. Extensive experiments have been conducted on seven datasets and the results have confirmed the superiority of the proposed model, including both prediction performance and running cost.

 {In {the} future work, we will study the following problems. First, the existing BroadCF model only considers rating information, and adding auxiliary data (e.g., reviews) can further improve the initial representation of users and items. Second, cross-domain recommendation has been widely studied to address the sparsity problem. However, {the} existing deep learning-based cross-domain recommendation algorithms also have the problem of facing higher computational complexity, i.e., consuming longer training time. In this context, it is an interesting research question to explore how broad learning system can be applied to cross-domain recommendation.}

\small{
\bibliographystyle{IEEEtran}
\bibliography{ref}

  \vskip -0.5in
\begin{IEEEbiography}
	[{\includegraphics[width=1in,height=1.25in,clip,keepaspectratio]{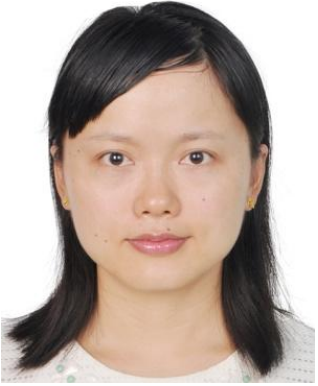}}]{Ling Huang} received her Ph.D. degree in computer science in 2020 from Sun Yat-sen University, Guangzhou. She joined South China Agricultural University in 2020 as an associate professor. She has published over 10 papers in international journals and conferences such as IEEE TCYB, IEEE TKDE, IEEE TNNLS, IEEE TII, ACM TIST, ACM TKDD, IEEE/ACM TCBB, Pattern Recognition, KDD, AAAI, IJCAI and ICDM. Her research interest is data mining.
\end{IEEEbiography}

\vskip -0.5in
\begin{IEEEbiography}
	[{\includegraphics[width=1in,height=1.25in,clip,keepaspectratio]{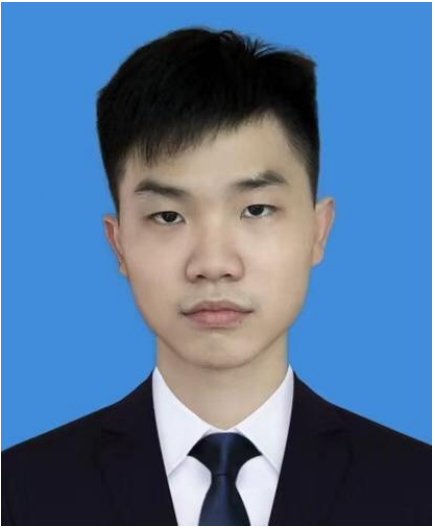}}]
	{Can-Rong Guan} received his B.E. degree in computer science in 2021 from South China Agricultural University. He is currently working toward the Master degree at South China Agricultural University. His research interest is data mining.
\end{IEEEbiography}

\vskip -0.5in
\begin{IEEEbiography}
	[{\includegraphics[width=1in,height=1.25in,clip,keepaspectratio]{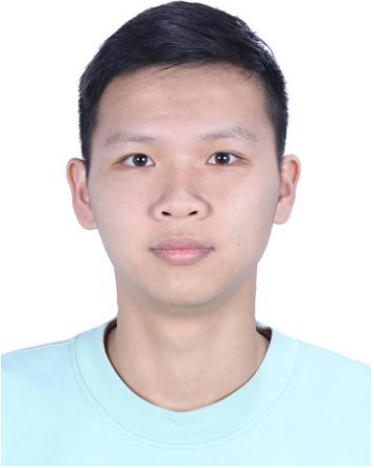}}]
	{Zhen-Wei Huang} received his B.E. degree in computer science in 2021 from South China Agricultural University. He is currently working toward the Master degree at South China Agricultural University. His research interest is data mining.
\end{IEEEbiography}

\vskip -0.5in
\begin{IEEEbiography}
	[{\includegraphics[width=1in,height=1.25in,clip,keepaspectratio]{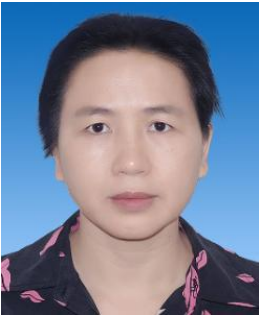}}]
	{Yuefang Gao} received the Ph.D. degree in computer science in 2009 from South China University of Technology, Guangzhou, China. She is a visiting scholar at the University of Sydney from March 2016 to April 2017. She joined South China Agricultural University in 2003, where she is currently an associate professor with School of Mathematics and Informatics. Her current research interests include computer vision and machine learning. She has published over 10 scientific papers in international journals and conferences such as TCSVT, Pattern Recognition and ACM MM.
\end{IEEEbiography}

\vskip -0.5in
\begin{IEEEbiography}
	[{\includegraphics[width=1in,height=1.25in,clip,keepaspectratio]{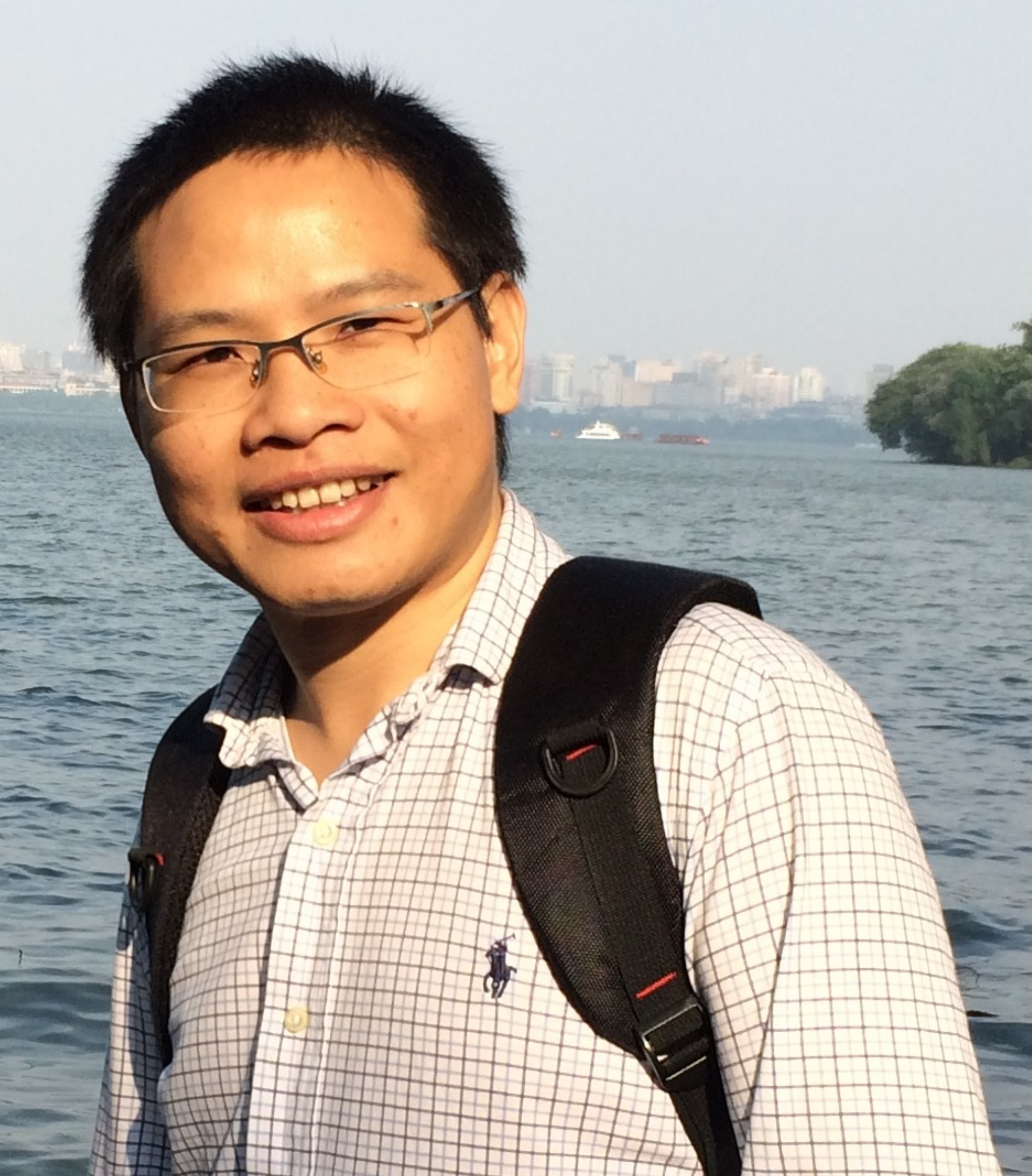}}]{Chang-Dong
		Wang} received the Ph.D. degree in computer
	science in 2013 from Sun Yat-sen University, Guangzhou,
	China. He was a visiting student with University of Illinois at Chicago from January 2012 to November
2012.
	He joined Sun Yat-sen University in
	2013, where he is currently an associate professor with School of Computer Science and Engineering.
	His current research interests include machine learning
	and data mining. He has published over 100 scientific papers in
	international journals and conferences such as IEEE TPAMI, IEEE TCYB, IEEE
	TKDE, IEEE TNNLS, KDD, AAAI, IJCAI and ACM MM. His ICDM 2010 paper won the Honorable Mention for Best
	Research Paper Awards. He won 2012 Microsoft Research Fellowship Nomination Award. He was awarded 2015 Chinese Association for Artificial Intelligence (CAAI) Outstanding Dissertation. His research works won 2018 First Prize of Guangdong Provincial Natural Science Award and 2020 Second Prize of Guangdong Provincial Natural Science Award respectively. He is an Associate Editor in Journal of Artificial Intelligence Research (JAIR). He is a Distinguished Member of China Computer Federation (CCF).
\end{IEEEbiography}

\vskip -0.5in
\begin{IEEEbiography}
	[{\includegraphics[width=1in,height=1.25in,clip,keepaspectratio]{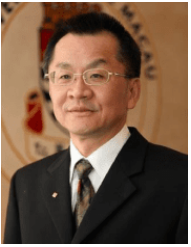}}]{C. L. Philip Chen} (Fellow, IEEE) received the M.S. degree in electrical engineering from the University of Michigan at Ann Arbor, Ann Arbor, MI, USA, in 1985, and the Ph.D. degree in electrical engineering from Purdue University, West Lafayette, IN, USA, in 1988.
	
	He is currently a Chair Professor and the Dean of the School of Computer Science and Engineering, South China University of Technology, Guangzhou, China. Being a Program Evaluator of the Accreditation Board of Engineering and Technology Education in the USA, for computer engineering, electrical engineering, and software engineering programs, he successfully architects the University of Macau’s Engineering and Computer Science programs receiving accreditations from Washington/Seoul Accord through the Hong Kong Institute of Engineers (HKIE), of which is considered as his utmost contribution in engineering/computer science education for Macau as the former Dean of the Faculty of Science and Technology. His current research interests include systems, cybernetics, and computational intelligence.
	
	Dr. Chen received the IEEE Norbert Wiener Award in 2018 for his contribution in systems and cybernetics, and machine learnings. He is also a 2018 Highly Cited Researcher in Computer Science by Clarivate Analytics. He was a recipient of the 2016 Outstanding Electrical and Computer Engineers Award from his alma mater, Purdue University. He was the IEEE Systems, Man, and Cybernetics Society President from 2012 to 2013. He was the Editor-in-Chief of the IEEE TRANSACTIONS ON SYSTEMS, MAN, AND CYBERNETICS: SYSTEMS, and the IEEE TRANSACTIONS ON CYBERNETICS. He is currently a Vice President of the Chinese Association of Automation (CAA). He is a member of the Academia Europaea, the European Academy of Sciences and Arts, and the International Academy of Systems and Cybernetics Science. He is a Fellow of IEEE, AAAS, IAPR, CAA, and HKIE.
\end{IEEEbiography}

\end{document}